\documentclass[apj]{emulateapj}
\usepackage{multirow}
\usepackage{graphics}
\usepackage{rotating}
\usepackage{enumitem}
\usepackage{amsmath}
\usepackage{txfonts}
\usepackage{hyperref}
\usepackage{setspace}
\usepackage{float}
\usepackage{tablefootnote}
\usepackage{footnote}
\usepackage{threeparttable}

\hypersetup{
			hyperfootnotes=true,
			colorlinks=true,
			linkcolor=blue,
			citecolor=blue,
			urlcolor=blue,
			breaklinks=false,
}

\shorttitle{Demographics of galactic bulges in SDSS}
\shortauthors{Kim et al.}

\begin{document}

\title{The Demographics of galactic bulges in the SDSS database}
\author{Keunho Kim\altaffilmark{1}, Sree Oh\altaffilmark{1}, Hyunjin Jeong\altaffilmark{2}, Alfonso Arag\'{o}n-Salamanca\altaffilmark{3}, Rory Smith\altaffilmark{1} and Sukyoung K. Yi\altaffilmark{1}}
\altaffiltext{1}{Department of Astronomy and Yonsei University Observatory, Yonsei University, Seoul 120-749, Korea; yi@yonsei.ac.kr} 
%\altaffiltext{2}{Yonsei University Observatory, Yonsei University, Seoul 120-749, Korea} 
\altaffiltext{2}{Korea Astronomy and Space Science Institute, Daejeon 305-348, Korea} 
\altaffiltext{3}{School of Physics and Astronomy, The University of Nottingham, University Park, Nottingham, NG7 2RD, UK}
%\altaffiltext{1}{}

% Forbidden Lines
\def\OI{[\mbox{O\,{\sc i}}]~$\lambda 6300$}
\def\OIII{[\mbox{O\,{\sc iii}}]~$\lambda 5007$}
\def\OIIIs{[\mbox{O\,{\sc iii}}]~$\lambda 4363$}
\def\OIIIab{[\mbox{O\,{\sc iii}}]$\lambda\lambda 4959,5007$}
\def\SIIab{[\mbox{S\,{\sc ii}}]~$\lambda\lambda 6717,6731$}
\def\SII{[\mbox{S\,{\sc ii}}]~$\lambda \lambda 6717,6731$}
\def\NII{[\mbox{N\,{\sc ii}}]~$\lambda 6584$}
\def\NIIb{[\mbox{N\,{\sc ii}}]~$\lambda 6584$}
\def\NIIa{[\mbox{N\,{\sc ii}}]~$\lambda 6548$}
\def\NI{[\mbox{N\,{\sc i}}]~$\lambda \lambda 5198,5200$}

% for table 1 
\def\OIIa{[\mbox{O{\sc ii}}]~$\lambda 3726$}
\def\OIIb{[\mbox{O{\sc ii}}]~$\lambda 3729$}
\def\NeIIIa{[\mbox{Ne{\sc iii}}]~$\lambda 3869$}
\def\NeIIIb{[\mbox{Ne{\sc iii}}]~$\lambda 3967$}
\def\OIIIa{[\mbox{O{\sc iii}}]~$\lambda 4959$}
\def\OIIIb{[\mbox{O{\sc iii}}]~$\lambda 5007$}
\def\HeII{{He{\sc ii}}~$\lambda 4686$}
\def\ArIVa{[\mbox{Ar{\sc iv}}]~$\lambda 4711$}
\def\ArIVb{[\mbox{Ar{\sc iv}}]~$\lambda 4740$}
\def\NIa{[\mbox{N{\sc i}}]~$\lambda 5198$}
\def\NIb{[\mbox{N{\sc i}}]~$\lambda 5200$}
\def\HeI{{He{\sc i}}~$\lambda 5876$}
\def\OI{[\mbox{O{\sc i}}]~$\lambda 6300$}
\def\OIb{[\mbox{O{\sc i}}]~$\lambda 6364$}
\def\SIIa{[\mbox{S{\sc ii}}]~$\lambda 6716$}
\def\SIIb{[\mbox{S{\sc ii}}]~$\lambda 6731$}
\def\ArIII{[\mbox{Ar{\sc iii}}]~$\lambda 7136$}

% Balmer lines
\def\Ha{{H$\alpha\,$}}
\def\Hb{{H$\beta\,$}}

% Line ratios
\def\NIIHa{[\mbox{N\,{\sc ii}}]/H$\alpha$}
\def\SIIHa{[\mbox{S\,{\sc ii}}]/H$\alpha$}
\def\OIHa{[\mbox{O\,{\sc i}}]/H$\alpha$}
\def\OIIIHb{[\mbox{O\,{\sc iii}}]/H$\beta$}

% Common terms
\def\Ebmv{E($B-V$)}
\def\LOIII{$L[\mbox{O\,{\sc iii}}]$}
\def\Ledd{${L/L_{\rm Edd}}$}
\def\LOIIIs4{$L[\mbox{O\,{\sc iii}}]$/$\sigma^4$}
\def\LOIIIMbh{$L[\mbox{O\,{\sc iii}}]$/$M_{\rm BH}$}
\def\Mbh{$M_{\rm BH}$}
\def\Msigma{$M_{\rm BH} - \sigma$}
\def\Ms{$M_{\rm *}$}
\def\Msun{$M_{\odot}$}
\def\Msunyr{$M_{\odot}yr^{-1}$}
\def\bt{$B/T\,$}
\def\btr{$B/T_{\rm r}\,$}

% Units
\def\ergs{$~\rm ergs^{-1}$}
\def\kms{${\rm km}~{\rm s}^{-1}$}
\newcommand{\cms}{\mbox{${\rm cm\;s^{-1}}$}}
\newcommand{\pccm}{\mbox{${\rm cm^{-3}}$}}
\newcommand{\sauron}{{\texttt {SAURON}}}
\newcommand{\oasis}{{\texttt {OASIS}}}
\newcommand{\HST}{{\it HST\/}}

\newcommand{\Vg}{$V_{\rm gas}$}
\newcommand{\Sg}{$\sigma_{\rm gas}$}
\newcommand{\eg}{e.g.,}
\newcommand{\ie}{i.e.,}

\newcommand{\gandalf}{{\texttt {gandalf}}}
\newcommand{\fracDeV}{{\texttt {FracDeV}}} 
\newcommand{\ppxf}{{\texttt {pPXF}}}

%==============================Abstract===========================================
\begin{abstract}
We present a new database of our two-dimensional bulge-disk decompositions for 14,233 galaxies drawn from SDSS DR12 in order to examine the properties of bulges residing in the local universe ($0.005 < z < 0.05$). We performed decompositions in $g$ and $r$ bands by utilizing the {\sc{galfit}} software. The bulge colors and bulge-to-total ratios are found to be sensitive to the details in the decomposition technique, and hence we hereby provide full details of our method. The $g-r$ colors of bulges derived are almost constantly red regardless of bulge size except for the bulges in the low bulge-to-total ratio galaxies ($B/T_{\rm r} \lesssim 0.3$). Bulges exhibit similar scaling relations to those followed by elliptical galaxies, but the bulges in galaxies with lower bulge-to-total ratios clearly show a gradually larger departure in slope from the elliptical galaxy sequence. The scatters around the scaling relations are also larger for the bulges in galaxies with lower bulge-to-total ratios. Both the departure in slopes and larger scatters are likely originated from the presence of young stars. The bulges in galaxies with low bulge-to-total ratios show signs of a frosting of young stars so substantial that their luminosity-weighted Balmer-line ages are as small as 1\,Gyr in some cases. While bulges seem largely similar in optical properties to elliptical galaxies, they do show clear and systematic departures as a function of bulge-to-total ratio. The stellar properties and perhaps associated formation processes of bulges seem much more diverse than those of elliptical galaxies.
\end{abstract}	


%==============================1. Introduction=====================================
\section{Introduction}
\label{sec:introduction}

The stellar bulges of disk galaxies are thought to be a complex mix of heterogeneous stellar populations, spanning a large range of age and chemical composition. Yet they show surprising similarities to elliptical galaxies.

In the modern cosmological framework, massive elliptical galaxies are thought to have formed through numerous mergers and interactions between smaller galaxies \citep{toom72, whit78}. Significant mergers are suspected to have happened more frequently in the earlier universe (say, $z>1$) and thus their stellar properties are rather uniformly old and metal-rich. Their dramatic evolution progressively changed their kinematic properties toward more chaotic orbits. These probably are the background for the tight scaling relations found for the stellar properties of elliptical galaxies, such as Faber-Jackson relation \citep{fabe76}, color-magnitude relation \citep{baum59, bowe92b}, black hole-bulge mass relation \citep{kim08,korm13}, and the fundamental plane \citep{dres87, djor87}. While the details are still debated, the community generally agrees on the early formation of massive elliptical galaxies \citep{cowi96,delu06,lee13}.

Would the similarities between bulges and elliptical galaxies then suggest similar formation history? This has been the main question in numerous previous investigations \citep{korm82, bend93, pele99, proc02, falc02, gado09}. Some studies indeed found evidence that bulges and ellipticals share very similar stellar properties based on stellar population analyses \citep{fish96,jabl96,falc02}. However, bulges may have wider varieties in their properties. For example, the presence of ``fake" bulges, also known as pseudobulges \citep{caro99, korm04}, and boxy/peanut bulges seems clear \citep{atha16}.

The relative size of bulge compared to the disk is known to correlate with galaxy morphology \citep{kent85, simi86, huds10}. However it is not trivial to explain why this is so. The similarity of the stellar bulge of a large Sa and a large elliptical galaxies may be easier to accept, but it is less clear why the small bulge of an Sd galaxy should share the same trends of stellar and kinematic properties as large ellipticals.

In this study, we investigate the properties of a large sample of bulges from the SDSS DR12 main galaxy sample, in the hope of answering some of these questions. We decompose the galaxies into bulge and disk components using the public tool {\sc{galfit}} \citep{peng10}. Comprehensive studies, with similar goals, using similar tools, conducted on similar samples, have been performed previously \citep{alle06,bens07,gado09,sima11,lack12,kelv12,meer15}. However, in some sense decomposition is often more art than science, and it is not all that clear how different studies derived their measurements. We perform our own decompositions and, unlike in many previous studies, we present the details of the input parameters and conditions in the decomposition routine so that future comparisons can be made. We describe a recipe for choosing the initial guess values of parameters, based on galaxy colour, and demonstrate its effectiveness.

Our bulge-to-total ratios correlate reasonably well with concentration index and Hubble type, but they correlate with S\'{e}rsic index more weakly, likely due to large uncertainties on S\'{e}rsic index measurements. We inspect the properties of bulges of galaxies as a function of bulge-to-total ratio. This effectively links spirals, lenticulars, and ellipticals into a long sequence of increasing $B/T$. We find that some bulges have very similar properties to ellipticals, while others differ in the slope, intercept, and scatter of scaling relations. We attempt to interpret these differences as a result of minor differences in the properties of stellar population, and to infer implications for the bulge formation.   

%This paper is organized as follows: sample selection for our study is described in Section \ref{sec:sample}; two-dimensional bulge - disk decomposition procedures and comparison of the results with the literature in Section \ref{sec:analysis}; results in Section \ref{sec:results}; discussion in Section \ref{sec:discussion}; and summary in Section \ref{sec:summary}. 
We adopt the $\Lambda$CDM cosmology of ($H_{\rm 0}$, $\Omega_{m}$, $\Omega_{\rm{\Lambda}}$) = (70${\rm km}~{\rm s}^{-1} {\rm Mpc}^{-1}$, 0.3, 0.7).

%===============================2. Sample Selection================================
\section{Sample Selection}
\label{sec:sample}

Our sample of galaxies is drawn from the Sloan Digital Sky Survey Data Release 12 \citep{york00, alam15}. We obtain basic observational parameters such as magnitude, angular size, b/a ratio of galaxies from CasJobs provided in DR12. We also employ the OSSY catalog \citep[][hereafter OSSY]{oh11} for spectroscopic information. This provides improved spectral measurements on the SDSS DR7 galaxies by utilizing GANDALF (Gas AND Absorption Line Fitting) and pPXF (penalized pixel-fitting) codes \citep{capp04, sarz06, oh11}.

We conduct all bulge and disk decompositions in the filters $g$ and $r$. 
%To achieve a more reliable decomposition, we perform our task only on close ($0.005<z<0.05$), apparently-large ($r_{\rm petro, r} > 10''$), and relatively face-on ($(b/a)_{\rm exp,r}>0.6$) galaxies. 
To achieve a more reliable decomposition, we perform our task only on close, apparently-large, and relatively face-on galaxies. 
After these selections 14,233 galaxies remained. 
We applied a volume limitation condition of $M_{\rm r} < -18.96$, which corresponds to the SDSS limiting apparent magnitude of $r < 17.77$ at $z=0.05$, to this database. The OSSY catalog provides spectroscopic measurements on 10,240 of these. The details of our initial selection criteria are given in Table 1.

%--------------------------Table 1; sample selection-------------------------------------------
\begin{table}
\centering
\begin{threeparttable}
\caption{Summary of initial sample selection}
\label{tab1}
\begin{tabular}{ll}
\hline \hline
Criterion & Explanation \\
\hline
$0.005 < z < 0.05$ & Redshift limit\\
 $r_{\rm petro,r}$\tnote{a} \, $> 10$'' & The minimum angular size\\ & of galaxy\\
 ${(b/a)_{\rm exp,r}}$\tnote{b} \, $> 0.6$ & Exclude severely edge-on galaxies\\
 ${r}\_{err_{\rm petro,r}}/{r_{\rm petro,r}} < 0.1$ & Relative error of angular size\\
 $M_{\rm r} < -18.96$ & Volume-limitation\\
 OSSY catalog & Cross-match for \\ & spectroscopic information\\
 \hline \hline \\
\end{tabular}
{\small
\begin{tablenotes}
\item[a] SDSS Petrosian radius in $r$ band
\item[b] SDSS exponential fit apparent axis ratio in $r$ band
\end{tablenotes}
}
\end{threeparttable}
\end{table}

%=========================3. Data Analaysis=====================================
\section{Data Analysis}
\label{sec:analysis}

%--------------------------Bulge-disk decompositions--------------------------------
\subsection{Two-dimensional Bulge - Disk Decompositions}

We performed two-dimensional bulge-disk decomposition in $g$ and $r$ bands by adopting the standard S\'{e}rsic and the exponential light profiles for the bulge and the disk, respectively \citep{sers68, free70}.
This combination has been suggested to be most suited to an automated and uniform decomposition on a large sample of galaxies \citep{meer13}.
We take the S\'{e}rsic index as a free parameter. The seeing of the SDSS (the median values of point spread function full width at half-maximum in $r$ band are $\sim 1.3''$) is probably not good enough to provide reliable values of S\'{e}rsic index; but we thought it would still be better to have the index free instead of fixed because it could be useful for characterizing the type of bulge (i.e. pseudobulges and classical bulges) at least on more trustworthy cases. 

The S\'{e}rsic light profile is described as follows \citep{sers68}:
\begin{eqnarray}
I(R) = I_{\rm e}{\rm exp}\left\{-b_{\rm n}[(R/R_{\rm e})^{1/n} -1]\right\}
\label{sersic-eq}
\end{eqnarray}
where $I(\rm R)$ is the surface brightness at a distance $R$ from the center of a galaxy, $R_{\rm e}$ is the effective radius which is defined such that half of the total light of a galaxy is contained within the effective radius, $I_{\rm e}$ is the surface brightness at the effective radius, and $n$ is the S\'{e}rsic index which shapes the light profile of a galaxy. We adopt the following expression of $b_{n}$ from \cite{capa89}:
\begin{eqnarray}
b_{n} \simeq 1.9992n - 0.3271.
\label{bn1}
\end{eqnarray}

To take advantage of having spatial information such as ellipticity and position angle, we used the two-dimensional decomposition technique \citep{byun95}. We utilize {\sc{galfit}} (version 3.0.5) for our two-dimensional galaxy decompositions \citep{peng02, peng10}. Basically, {\sc{galfit}} adopts the Levenberg-Marquardt algorithm which is based on the least square minimization technique. Further details on the fitting algorithm and usage of {\sc{galfit}} are described in \cite{peng02,peng10}. A comparison study of {\sc{galfit}} and another widely-used piece of software {\sc{gim2d}} is conducted in \cite{haus07}.

The {\sc{galfit}} software requires four kinds of input image files for decomposition: 1) observation image, 2) background noise image (a.k.a. $\sigma$ image). 3) PSF image, and 4) masking image to remove neighboring galaxies and foreground stars. The observation images of sample galaxies are drawn from SDSS DR12 Science Archive Server; specifically, we use the calibrated and sky-subtracted frame fits images. The $\sigma$ images are internally generated in {\sc{galfit}} with essential information such as exposure time, gain, readout noise, and the number of combined fits from the fits header. The PSF images are generated by utilizing SDSS \textbf{readAtlasImages-5$\_$4$\_$11}. The \textbf{SExtractor version 2.8.6} \citep{bert96} was used to identify all objects in given observational images and to generate masking images for our sample galaxies.

%--------------------------Table 2; Input initial guess values and fitting constraint range-----------------------------
\begin{table*}
\centering
\begin{threeparttable}
\caption{Input initial guesses and constraints range of fitting parameters}
\label{tab2}    
\begin{tabular}{l c c c} \hline \hline

Parameter	&	\multicolumn{2}{c}{Initial guess}	&	Constraint Range	\\	  
\hline

	&	$(g-r)_{\rm galaxy}$\tnote{a} $ > 0.65$	&	$(g-r)_{\rm galaxy} \le 0.65$	&	\\
\hline
Bulge    &		&		&		\\ 
\hline

x / y center	&	\multicolumn{2}{c}{$R.A.$\tnote{b} and $Dec$\tnote{c}}	&	$R.A.$ $\pm 5$ pix and $Dec$ $\pm 5$ pix	\\

Magnitude	& Petromag$_{\rm r}$\tnote{d} & Petromag$_{\rm r} + 3$ & Petromag$_{\rm r} \pm 5$ \\

Effective radius	&	10 pix	&	5 pix	&	3 pix to ${r_{\rm petro,r} - 5}$ pix	\\

S\'{e}rsic index	&	4	&	1	&	1 to 8	\\

Apparent axis ratio	& \multicolumn{2}{c}{$(b/a)_{\rm dev,r}$\tnote{e}}	&	0.3 to 1.0	\\
Position Angle	& \multicolumn{2}{c}{$P.A._{\rm dev,r}$\tnote{f}}	&		\\
\hline
Disk    &		&		&		\\ 
\hline
x / y center	&	\multicolumn{2}{c}{Fixed to be same as those of bulge}	&	\\

Magnitude	& Petromag$_{\rm r} + 3$ & Petromag$_{\rm r}$ & Petromag$_{\rm r} \pm 5$ \\

Scale length	&	5 pix	&	12 pix	&	5 pix to ${r_{\rm petro,r} + 5}$ pix	\\

Inclination	& \multicolumn{2}{c}{$b/a_{\rm exp,r}$\tnote{g}}	&	$b/a_{\rm exp,r} \pm 0.1$	\\
Position Angle	& \multicolumn{2}{c}{$P.A._{\rm exp,r}$\tnote{h}}	&	$P.A._{\rm exp,r} \pm 20$ degrees  \\

\hline 
\hline

\end{tabular}
{\small
\begin{tablenotes}
\item[a] $g-r$ galaxy color
\item[b] SDSS right accession of galaxy
\item[c] SDSS declination of galaxy
\item[d] SDSS Petrosian magnitude in $r$ band
\item[e] de Vaucouleurs fit apparent axis ratio in $r$ band
\item[f] de Vaucouleurs fit position angle in $r$ band
\item[g] Exponential fit apparent axis ratio in $r$ band
\item[h] Exponential fit position angle in $r$ band
\end{tablenotes}
}
\end{threeparttable}
\end{table*}

Inserting these four input images into {\sc{galfit}}, we performed bulge - disk decompositions. We chose the initial guesses of bulge and disk components based on the SDSS observation quantities such as $R.A.$, $Dec$, $Petromag\_{\rm r}$, $Petromag\_{\rm g}$, $Petrorad\_{\rm r}$, $deVAB\_{\rm r}$, $expAB\_{\rm r}$, $deVPhi\_{\rm r}$, and $expPhi\_{\rm r}$. As mentioned in other previous studies, the `mathematical' best fit does not always guarantee a solution that is physical \citep{peng10,meer15}. We try to tackle this limitation by adopting initial guesses and fitting constraint ranges that are astrophysically motivated. There are numerous studies which discuss effects of initial guesses on the results of fits \citep{broe97,maca03,gado09,wein09,fish10,fern14}. After some experimentation, we found that the $(g-r)$ color is the most effective galaxy property for deciding initial guesses. This works well simply because $(g-r)$ is a reasonable morphology indicator.\footnote{The referee pointed out that $(g-i)$ can be similarly effective. We confirm that a $g-i > 0.9$ criterion leads to a virtually identical result.} Our initial guesses and constraint ranges are given in Table \ref{tab2}.
 
%----------Figure 1-------Examples of B + D decompositions
\begin{figure}
\centering
\includegraphics[width=0.5\textwidth]{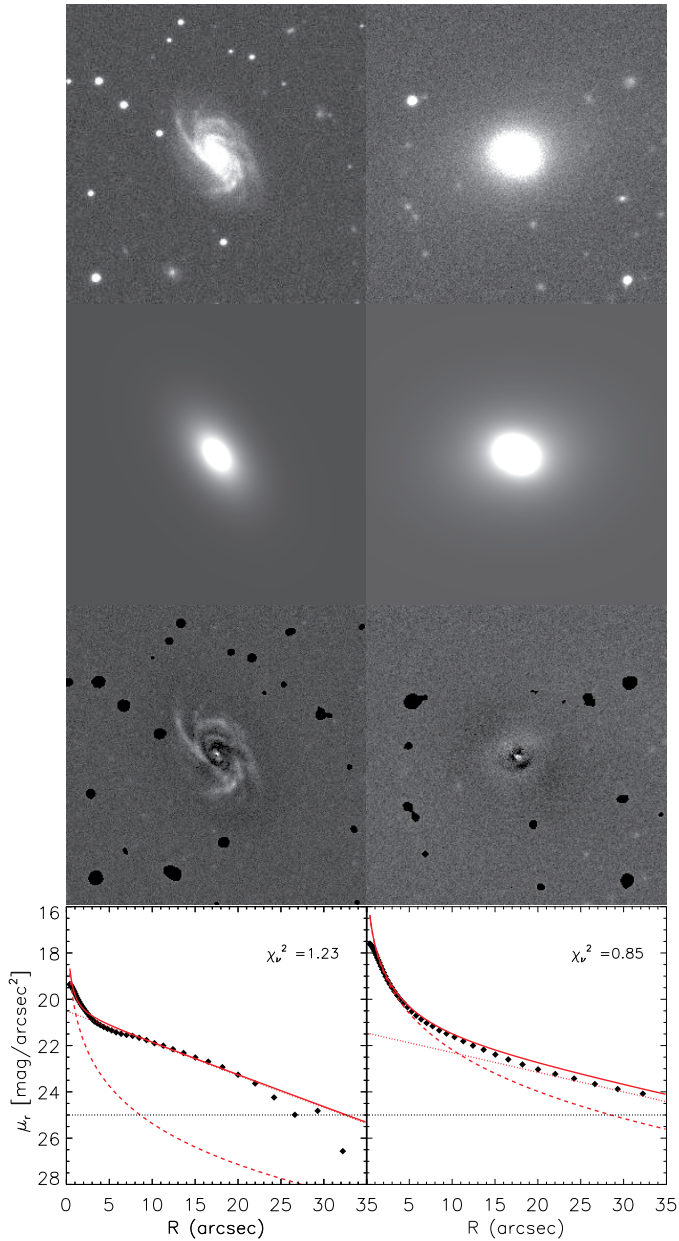}
\caption{Two examples of bulge-disk decompositions. From top to bottom, $r$-band observation image, {\sc {galfit}} model image, model-subtracted residual image, and 1-D radial profile are displayed. The black diamond is the median value of observed surface brightness at a given radius, measured from {\sc {IRAF}} ellipse task. The red dashed and dotted lines indicate the radial profile of bulge and disk components, respectively. The red solid line is the sum of surface brightness of bulge and disk components. The horizontal line shows the limiting surface brightness of SDSS. The reduced $\chi^{2}$ for each galaxy is marked at the top-right in the 1-D radial profile panel.}
\label{fig1}
\end{figure}

%----------Figure 2----------Chi^2 and Sersic index ditribution
\begin{figure}
\centering
\includegraphics[width=0.5\textwidth]{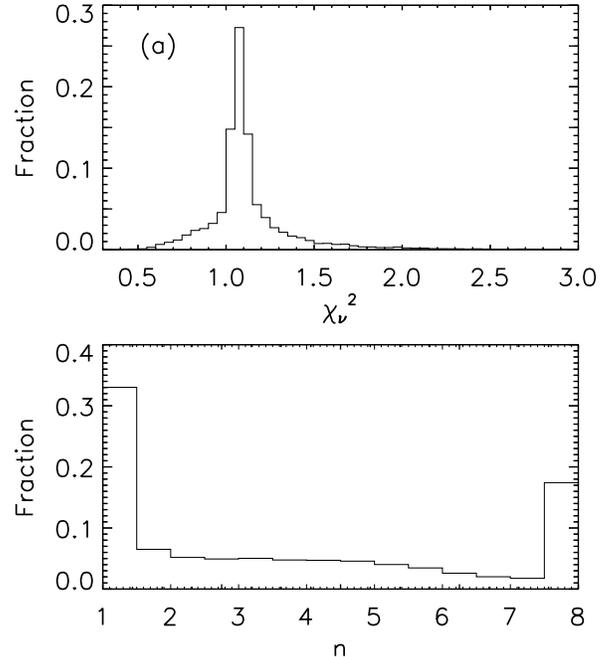}
\caption{The reduced $\chi^{2}$ {\em(panel a)} and bulge S\'{e}rsic index {\em(panel b)} distribution of our decompositions.}
\label{fig2}
\end{figure}

Two examples of the resulting bulge-disk decompositions are shown in Fig. \ref{fig1}. 
We first performed the decomposition in $r$ band as described above. We then applied this result to the $g$-band decomposition by fixing all the fitting parameters except for the positions of centers and the bulge and disk magnitudes. This ``simultaneous fitting'' technique has been reported to minimize the errors of fits \citep{sima02,sima11} and is widely used \citep[e.g.,][]{sima11, lack12, meer15}. We confirm this through a test on sample galaxies. A drawback of this technique, on the other hand, is that color gradients cannot be measured using this method. Hence, we do not discuss color gradients further in this study. For the bulge fit, we explored the whole range of $n$ down to the index of exponential disks (1.0).

We note that there are some recent findings (e.g., \cite{erwi15}) on low mass hot classical-like bulges having $n < 1$, so that some bulges in our sample galaxies might possibly have the intrinsic S\'{e}rsic index less than 1\footnote{This was kindly pointed out by the referee.}. However, following the classical definition of a bulge component as a centrally compact stellar region of a galaxy, we adopted a bulge S\'{e}rsic index range of $1 \leq n \leq 8$. Therefore, our bulge component has a steeper light profile than the exponential disk component. We confirm that permitting a S\'{e}rsic index range with $n$ lower than 1 has negligible impact on our results throughout the paper.

%\footnote{\textbf{We note that there are some recent findings (e.g., \cite{erwi15}) on low mass hot classical-like bulges having $n < 1$, so that some bulges in our sample galaxies might possibly have the intrinsic S\'{e}rsic index less than 1. However, following the classical definition of a bulge component as a centrally compact stellar region of a galaxy, we adopted a bulge S\'{e}rsic index range of $1 \leq n \leq 8$. Therefore, our bulge component has a steeper light profile than the exponential disk component. We confirm that permitting a S\'{e}rsic index range with $n$ lower than 1 has negligible impact on our results throughout the paper.}}. 
\label{sec:decomposition}

%------------------------------------Quality Estimation of Fit-----------------------------------------
\subsection{Quality Estimation of Fits}

Here we discuss the quality of our fits based on the $\chi^{2}$ measurement. The value of $\chi^{2}$ reflects the goodness of fits, but a small value does not always guarantee a physically-sound fit \citep{meer15}. The reduced chi-squared ($\chi^{2}_{\nu}$) distribution of our decompositions is shown in Fig. \ref{fig2}-(a). The peak value is around 1.1, and the distribution shows a gradual decrease from the peak on both sides. We inspected some galaxies with $\chi^{2} _{\nu}< 0.6$ or $\chi^{2} _{\nu}> 3$: that is, their color-composite images, model and residual images of decompositions, and one-dimensional radial profiles. The majority of galaxies having a small $\chi^{2}_{\nu}$ are red early-type galaxies. This is caused by the fact that their relatively simple structure is fitted tightly already with one component (bulge) and an additional component (i.e., disk) is largely unnecessary. On the other hand, those with a large value of $\chi^{2}_{\nu}$ are complex systems that are poorly fitted by just a disk and bulge mix. Examples include interacting galaxies and late-type galaxies with prominent features such as spiral arms, rings, or bars. Among the poor-$\chi^{2} $ galaxies were some featureless galaxies, those with strong dust extinction, and those where SDSS pointing fails to find the center of the galaxies. We removed 264 galaxies which have $\chi^{2}_{\nu} \geq 2$ as we consider them poor fits, leaving 9,976 galaxies behind.

Fig. \ref{fig2}-(b) shows the distribution of the S\'{e}rsic index $n$ of {\em bulges}. More than 30 percent of bulges have low values ($1.0 \leq n \leq 1.5$). This roughly agrees with the late-type galaxy fraction found in the local universe \citep{oh13, khim15}. There is another peak at $7.5 \leq n \leq 8.0$. Their mean value of $\chi^2_{\nu}$ is 1.24 and not very different from the rest of the sample. Their mean $B/T_{\rm r}$ is 0.15. We inspected the images of these objects and found that the fit is not particularly poorer than the rest of the sample. Instead we found a point-like source in the galaxy center. Indeed, \citet{wein09} pointed this out earlier and used a third point-like component at the center in addition to disk and bulge. In our study, we adopted the two-component decomposition technique and thus the presence of this peak may at least partly be a result of the shortcoming of our two-component approach. We will discuss the reliability of $n$ measurements in Section 4.2 in more detail. 
\label{sec:quality estimation}

%------------------------Comparison with the Literature---------------------------
\subsection{Comparison with the Literature}

We compare our results with \citet[][hereafter S11]{sima11} and \citet[][hereafter M15]{meer15} because they too used the same fitting functions as ours: a free S\'{e}rsic function for the bulge component and an exponential disk. Cross-matching with S11 and M15 resulted in 11,384 and 11,345 galaxies, respectively.

%----------Figure 3----Comparions of B/T with S11 and M15
\begin{figure}
\centering
\includegraphics[width=0.5\textwidth]{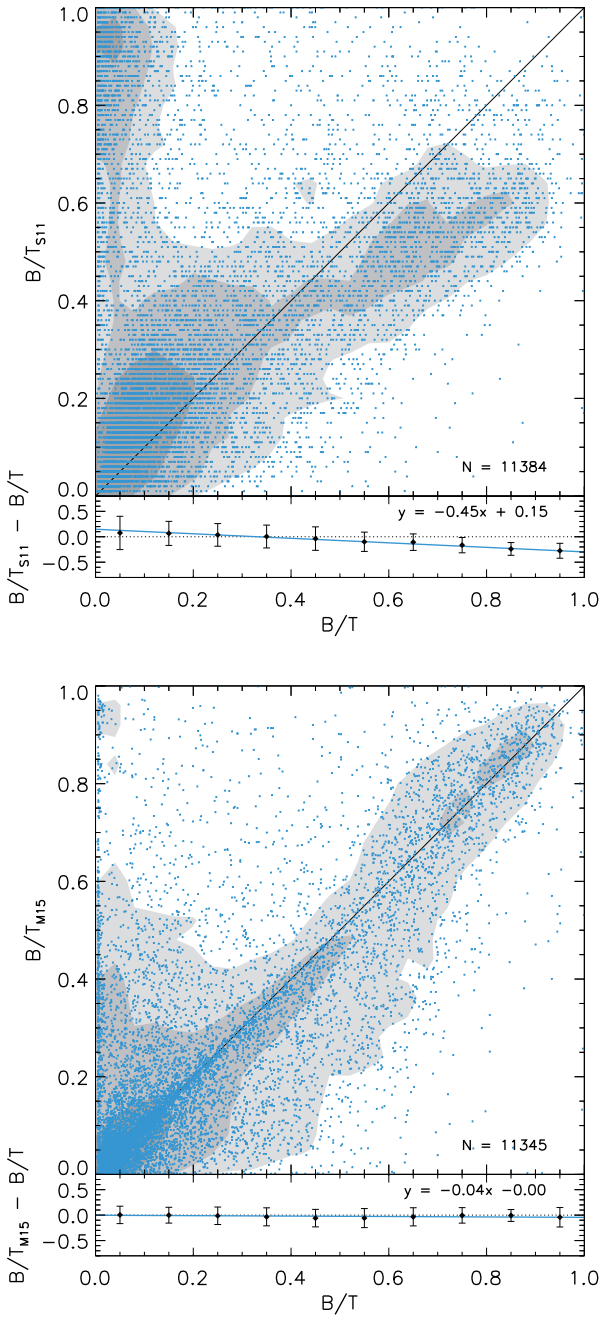}
\caption{Comparison of $B/T$ in the $r$ band between this study and those of S11 (top) and M15 (bottom). The numbers of galaxies cross-matched are given. The gray shaded-contours indicate the galaxy distribution at $0.5$, $1$, and $1.5 \sigma$ levels. The black solid line is the one-to-one reference line. The sub-panels of the two comparisons show the difference in $B/T_{\rm r}$. The black diamonds show the median values, and the corresponding error bars show the $1\sigma$ standard deviation. The the blue solid lines and the equations given are the linear fits to the median values. The dotted lines in the sub-panels are a reference of zero difference. }
\label{fig3}
\end{figure}

Fig.\ref{fig3} shows the comparison between the databases in terms of $B/T$, the most important parameter we aim to derive in this study. The agreement is substantially better with M15 (bottom panel) which is at least partially due to the fact that we use the same decomposition software {\sc{galfit}} as M15, while S11 used {\sc{gim2d}}, and it has been shown that different fitting algorithms often indicate different values of structural parameters \citep{haus07,gado09,sima11,meer15}. 

Agreement is generally better for bulge-dominant galaxies in both cases. The poorest agreement is found on small $B/T$ galaxies and between S11 and us. For example, there are many galaxies for which S11 derived a much higher value of $B/T$ than us (the vertical band of points at $B/T_{\rm r} < 0.1$ in the top panel of Fig. \ref{fig3}). We inspected their images and fits individually and found that our fits were generally more trustworthy. But, more importantly, we would like to emphasize the fact that it is very difficult to measure $B/T$ on disk-dominant galaxies using the techniques that are widely used today especially on the observational data of SDSS quality. We could not pin down the source of differences between studies because details of the parameters used for decompositions in previous databases are not explicitly given. It is our hope that giving full details of our fitting procedure and parameters will make future comparison more feasible. 
\label{sec:comparions with the Literature}

%==================================4. Results==================================
\section{Results}
\label{sec:results}
%---------------------------Section 4.1--------------------------------
\subsection{Optical Colors of Bulges}
\label{sec:Optical Colors of Bulges}

\begin{figure}
\centering
\includegraphics[width=0.5\textwidth]{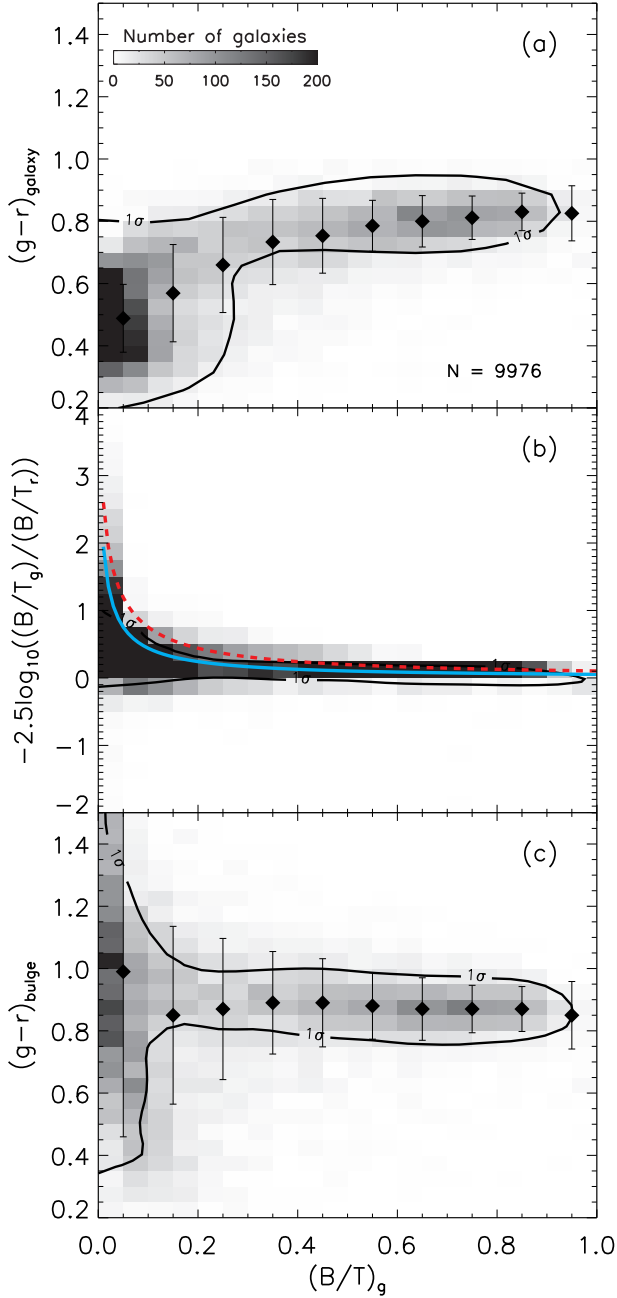}
\caption{The Hess diagram for galaxy and bulge colors vs $B/T_{\rm g}$. 
{\em(panel a)} Galaxy colors show a general trend of being redder with increasing $B/T$. {\em(panel b)} The difference in $B/T$ in $g$ and $r$ vs. $B/T_{\rm g}$. The blue solid and red dashed lines indicate the cases of $B/T_{\rm r} \, - \, B/T_{\rm g} = 0.05$ and 0.1, respectively.  
{\em(panel c)} The bulge colors derived; i.e., $(a) + (b)$. The solid line shows the 1$\sigma$ contour. The median values and standard deviations are given. The total number of galaxies used ($N$) is marked on bottom-right in {\em(panel a)}.}
\label{fig4}
\end{figure}

The color of a galaxy provides information on the age, metallicity of constituent stellar populations and internal dust extinction. It is thus interesting to check if the colors of bulges we derive here show any trend and to see how such trends can be interpreted. An accurate derivation of the colors of bulges requires reliable measurements of $B/T$ in the associated bands. We present bulge colors against $B/T$ in Fig. \ref{fig4}-(c). By and large, the bulge colors are measured to be uniformly red within errors. 

The bulge colors of disk-dominated ($B/T_{\rm r} \lesssim 0.1$) galaxies are noteworthy. Their median colors are similar to the rest of the sample but with a much larger scatter. There are quite a few bulges that are unusually blue (e.g., $g-r < 0.8$) for a bulge. Their composite images show blue colors consistently, and if our decomposition is reliable, it would mean that some bulges of the low $B/T$ galaxies are indeed of notably different stellar properties from the rest of the sample. 

There also are a large number of excessively red bulges at $B/T_{\rm r} \lesssim 0.1$. Note that the typical color of the largest $B/T$ bulges is $g-r \approx 0.8$ (Fig. \ref{fig4}-(c)). It is consistent with the mean color of elliptical galaxies that contain a negligible amount of dust. If bulges are similarly devoid of dust, their extremely red colors in the lowest $B/T$ bin are difficult to understand. Even if we adopt an unexpectedly large value of dust extinction of $E(B-V)=0.2$, a population similarly old and metal-rich to the bulges in the large $B/T$ galaxies only become as red as $g-r \approx 1.2$. In Fig. \ref{fig4}-(c), there is a long tail toward red colors, extending vertically beyond the limits of the figure, with some bulge colors derived to be as red as $g-r > 2.0$!. These bulge colors do not appear astrophysically plausible. This has been noted previously by other studies \citep{mend14,fern14}. 

However excessively red bulge colors derived from decomposition can be explained. The excessively red colors of bulges are likely caused by a methodological limitation. Importantly, the logarithmic ratio of $B/T$ values in the associated bands is employed in the derivation of bulge color from galaxy color. The bulge color, $(g-r)_{\rm bulge}$, is calculated as follows:

%---------------------Eq 3: the anti-correlation-------------------------------------------
\begin{eqnarray}
(g-r)_{\rm bulge} &=& m_{\rm g,b} - m_{\rm r,b}\nonumber \\
                         &=& m_{\rm g,g} - m_{\rm r,g} - 2.5\,{\rm log}_{10}((B/T_{\rm g})/(B/T_{\rm r}))
\label{Eq: anti_correl_eq}
\end{eqnarray}
where $m_{\rm g,b}$ and $m_{\rm r,b}$ are $g$ and $r$-band magnitudes of bulge, $m_{\rm g,g}$ and $m_{\rm r,g}$ are $g$ and $r$-band magnitudes of galaxy, and $B/T_{\rm g}$ and $B/T_{\rm r}$ are $g$ and $r$-band bulge to total ratio, respectively. As shown in Equation 3, bulge color is derived from galaxy color (Fig. \ref{fig4}-(a)) and the ratio between the values of $B/T$ in $g$ and $r$ bands (Fig. \ref{fig4}-(b)). The large ratios of the small $B/T$ galaxies in panel $b$ are the cause of the red tail in the small $B/T$ galaxies in panel $c$. 

%----------Figure 4-----g-r bulge color derivation process

%----------Figure 5---B/T_g vs B/T_r
\begin{figure}
\centering
\includegraphics[width=0.5\textwidth]{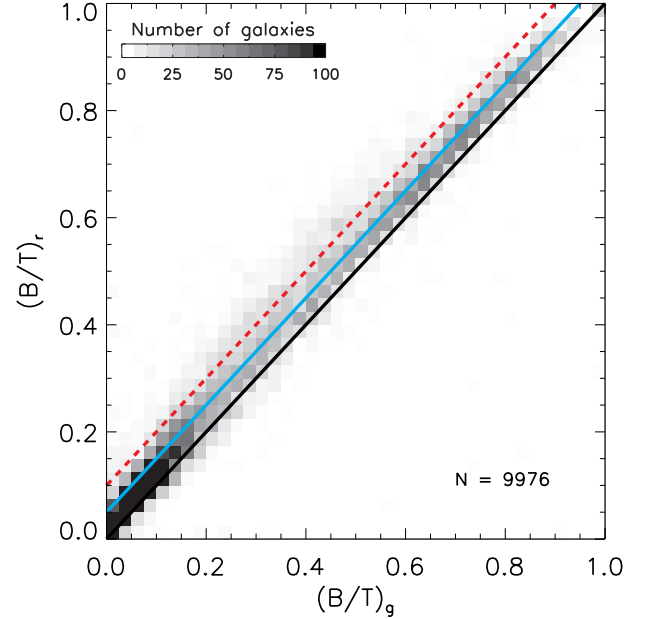}
\caption{Bulge-to-total ratios in $g$ and $r$ from this study. The black solid line is the one-to-one reference line. The blue solid line and the red dashed line are the same as in Fig. \ref{fig4}-($b$).}
\label{fig5}
\end{figure}

Figure \ref{fig5} shows the comparison between the $B/T$ values measured in $g$ and $r$ bands. The agreement looks good from a cursory inspection, but the ratio between them, as appears in Equation 3, can be dramatically large especially in small $B/T$ galaxies. For example, if we assume that the typical difference in $B/T$ between $g$ and $r$ bands is $\delta \approx 0.05$, that translates to $\Delta (g-r) = 0.4$ and 0.07 for the galaxies of $(B/T)_{\rm g} = 0.1$ and 0.8, respectively. More dramatic cases ($\delta > 0.05$) can be easily found simply by looking further into the details of Figure \ref{fig5}. In conclusion, the excessively red colors of bulges derived in our decompositions and other studies probably do not reflect the real astrophysical properties of bulges but instead a shortcoming of the methodology which is currently widespread. Small differences in $B/T$ measured in different wavelengths can result in significant bulge color reddening as $B/T$ becomes small. We have thus decided to remove 3,956 galaxies with $B/T_{\rm r} \leq 0.1$ out of 9,976, leaving only 6,020 galaxies for our further analysis.

%---------------------------Bulge to total ratio and galaxy morphology------------------------------
\subsection{Bulge to total ratio and galaxy morphology}
\label{sec:Bulge to total ratio}

\begin{figure*}[ht]
\centering
\includegraphics[width=1\textwidth]{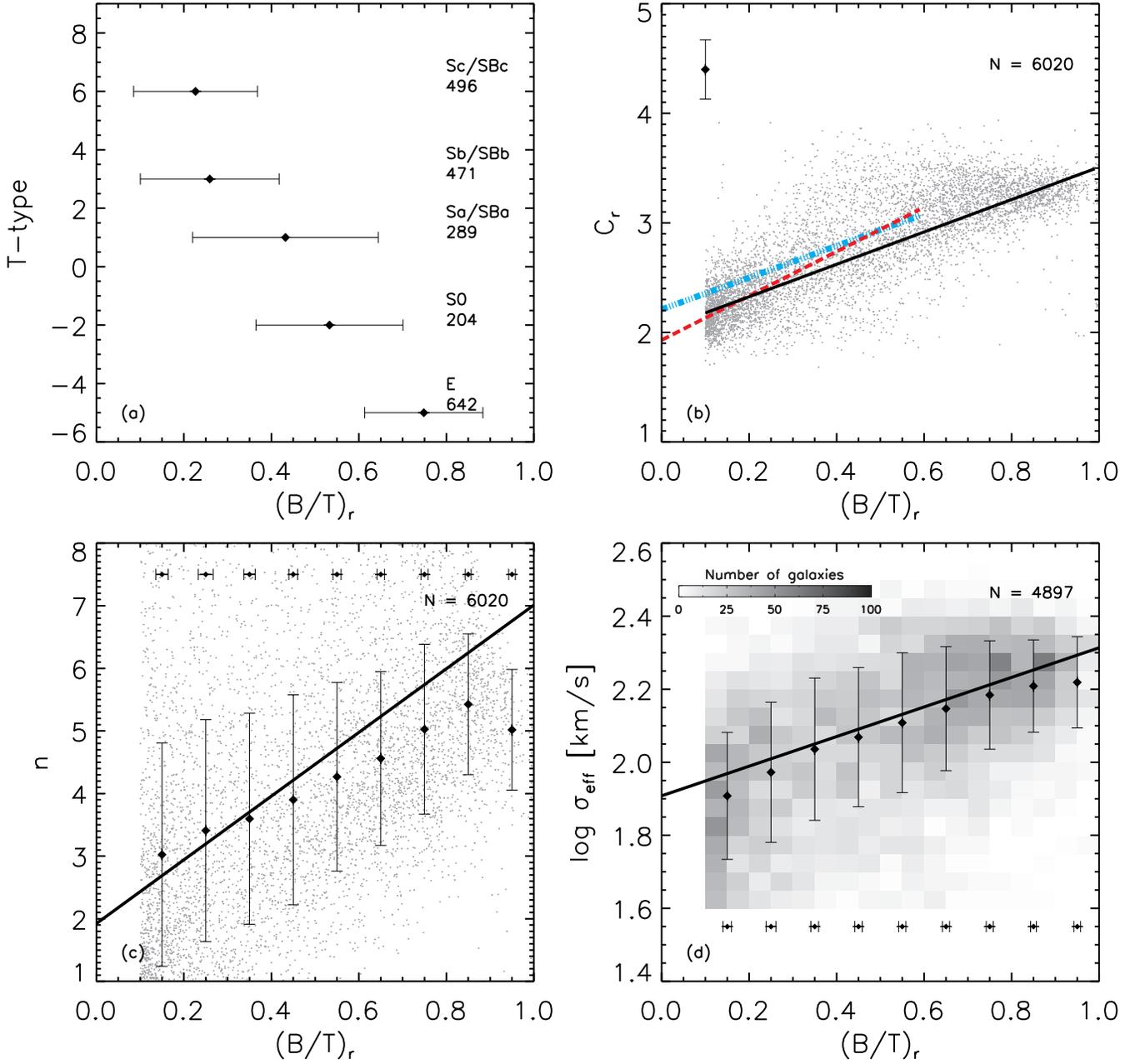}
\caption{Correlation between $B/T$ and other morphology indicators.
{\em(panel a)} 
Hubble type. Hubble t-type is adopted from \cite{khim15}. The number of galaxies of each t-type is marked on top. The mean and standard deviation are shown as black diamonds with error bars. 
{\em (panel b)}
Concentration index. The black solid line is the linear fit to our sample galaxies for the range of $0.1 \leq B/T_{\rm r} \leq 1.0$. The red dashed and blue dot-dashed lines are the linear fits of \cite{gado09} and \cite{lack12} for their galaxies, respectively. Note that their fitting ranges were different: $0.0 \leq B/T_{\rm r} < 0.6$. The standard deviation of our fit residuals is marked on top-left. The number of galaxies is also shown. 
{\em (panel c)}
The S\'{e}rsic indices of $bulges$. The mean and standard deviation are shown. The solid line is the linear fit to the whole sample. The Pearson correlation coefficient is only 0.30. The number of galaxies is marked on top-right of the panel. The error bars on top of the panel shows the error of B/T in each range based on the errors of our decomposed bulge and disk magnitudes from {\sc{galfit}} which are typically underestimated \citep{zhao15}.
{\em (panel d)}
Central velocity dispersion $\sigma_{\rm eff}$. The mean and standard deviation are shown. The solid line is the linear fit to the whole sample. The Pearson correlation coefficient is 0.58. The error bars at bottom of the panel shows the error of B/T in each range based on the errors of our decomposed bulge and disk magnitudes from {\sc{galfit}}.}
\label{fig6}
\end{figure*}

The relative size of the bulge component in a galaxy has been known to correlate with the morphology of its host galaxy in the sense that the value of $B/T$ becomes larger as galaxy morphology moves from late to early types \citep{hubb26, hubb36, kent85, simi86, huds10}. As a sanity check, we compare our $B/T$ values against galaxy morphology and other morphology indicators (concentration index, the S\'{e}rsic index, and velocity dispersion) in Fig. \ref{fig6}. The concentration index used here is defined as $C_{\rm r} \equiv {\rm PetroR90/PetroR50}$ adopting SDSS Petrosian radii. Galaxy morphology information is from \cite{khim15} who performed visual morphology classification on nearby ($0.025 < z <0.044$) SDSS DR7 galaxies. The S\'{e}rsic indices shown are those derived from our decomposition.

%----------Figure 6----B/T vs Cr, T-type, Sersic index, and velocity dispersion

The $B/T$ ratio indeed correlates well with morphology (Fig. \ref{fig6}, panel a). Although there is a fairly large scatter, the trend is clear. We note that the $B/T$ of late type spirals (e.g. at $T=6$) is somewhat overestimated because we excluded low $B/T$ galaxies ($B/T_{\rm r} \le 0.1$).

The correlation with concentration index (panel b) is also reasonably clear (the Pearson correlation coefficient is $\sim 0.79$ for galaxies with $0.1 <  B/T_{\rm r} \leq 1.0$), as recently confirmed on similar samples \citep{gado09, lack12}. We derived a linear fit (solid line) to the sample with $0.1 < B/T_{\rm r} \leq 1.0$ for which decomposition seems more reliable and the resulting linear fit is as follows:
%-------------------------Eq. 4------------------------------------------
\begin{eqnarray}
C_{\rm r} &=& (1.48 \pm 0.03) \ B/T_{\rm r} + 2.03 \pm 0.03.
\label{Eq: Cr vs BT_r}
\end{eqnarray}
For comparison, the fits of earlier studies \citep{gado09, lack12} are also shown in the Figure. 

The panel $c$ shows the comparison against the S\'{e}rsic index $n$ of the bulge component. The solid line shows the least square linear fit. The Pearson correlation coefficient is 0.30 and thus the correlation is not strong, but the trend is as expected. It is very difficult to accurately measure the S\'{e}rsic index from {\em two-component} decomposition on {\em SDSS-quality} images.
For example, the poor point spread function of the SDSS data causes $n$ to be measured lower than true, while lacking a point-like source at the galactic center in the decomposition procedure tends to result in a higher value of $n$ than true.  
%, which probably is the main cause for the large scatter {\bf{(PSF causes offset to lower values, not scatter? If so, high values not due to PSF effects)}}. 
But it might also imply that a larger bulge does not strictly mean a larger value of $n$ of the $bulge$.

Morphology is known to correlate with velocity dispersion, too \citep[e.g.,][]{oh13, khim15}. Naturally, a positive correlation between $B/T$ and velocity dispersion has been reported \citep[e.g.,][]{huds10}. We show our result in panel $d$. We employ the velocity dispersion from the OSSY catalog and apply the aperture correction for the bulge circular effective radius following \cite{capp06} which derived the aperture-correction relation based on a luminosity-weighted spectra. We chose only galaxies with $40 < \sigma_{\rm eff} < 400 \ {\rm km}~{\rm s}^{-1}$ and ${\rm error} (\sigma_{\rm eff})/\sigma_{\rm eff} < 0.5$ to ensure a good quality of velocity dispersion measurement. This cut in velocity dispersion and associated error removes 19\% (1123 out of 6020 galaxies) of our sample. The removal fraction is larger for galaxies with smaller values of $B/T$: 39\% for $0.1 < B/T \leq 0.3$, 16\% for $0.3 < B/T \leq 0.5$, 4\% for $0.5 < B/T \leq 0.7$, 0.9\% for $0.7 < B/T$. A positive but weak correlation between the two parameters is found, which agrees well with the result on the cluster galaxy sample of \cite{huds10}.

Compared to classical bulges, pseudobulges are in general said to have lower values of velocity dispersion \citep{korm04,korm15}. Recently, \citet{fabr12} suggested that they generally have ${\rm log}\ \sigma \lesssim 2.0$. This corresponds to $B/T_{\rm r} \lesssim 0.4$ based on the fit found from our sample. This issue will be revisited in Section 4.4.

The $B/T$ ratios derived from our decomposition appear to be sensible. This indirectly demonstrates that our scheme for choosing the initial guess parameters, and their boundaries for decomposition, was effective. All in all, it indeed seems possible to describe the whole Hubble sequence as a single sequence of $B/T$.

%---------------------------Color-Magnitue Diagram--------------------------------
\subsection{Color-Magnitude Diagram}
\label{cmd analysis}

The Color-Magnitude Diagram (CMD) has been widely used to study the formation and evolution of galaxies \citep[e.g.,][]{sand78, bowe92b, stra01, blan03, driv06}. We now inspect the bulge properties in the CMD in the hope of finding some clues to their formation.

\begin{figure}
\centering
\includegraphics[width=0.5\textwidth]{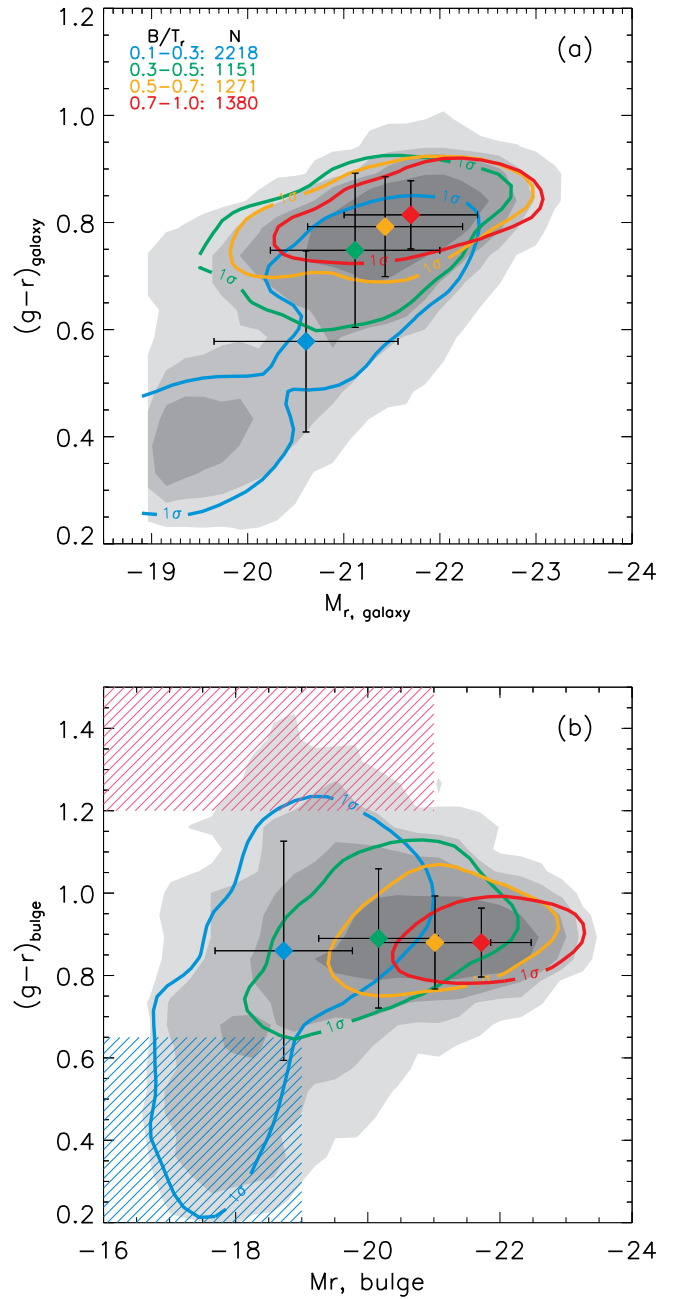}
\caption{The color-magnitude diagram with $B/T_{\rm r}$ binning for galaxies {\em(panel a)} and bulges {\em(panel b)}. The gray-shades show the 0.5, 1, 1.5, and 2$\sigma$ contours for the whole sample. The blue, green, orange, and red solid lines represent the 1$\sigma$ contours binned as noted. The number of galaxies in each $B/T$ range is marked on top-left of {\em panel a}. The median value and standard deviation are also given. The red shaded area in {\em panel b} indicates the region in which unphysically-red bulge colors are derived, as described in Section \ref{sec:Optical Colors of Bulges}. The blue shaded area indicates the region in which star-forming bulges are mainly identified. We note that the additional sample selection criterion as mentioned in Section \ref{sec:Bulge to total ratio} based on velocity dispersion removes a considerable number of the galaxies found mainly in the blue cloud.}
\label{fig7}
\end{figure}

Fig. \ref{fig7} shows the CMD of our sample. Panel $a$ shows the {\em galaxy} properties.
Contours present the $1\sigma$ distribution of galaxies split into groups based on $B/T_{\rm r}$. Note that the 1$\sigma$ contours gradually shift from the blue cloud to the red sequence with increasing $B/T$. Wide varieties of $B/T$ are present in the red sequence region. On the other hand, blue cloud galaxies ($g-r \lesssim 0.6$) almost exclusively have low values of $B/T$.

%----------Figure 7------Galaxy and bulge CMD with B/T marked

Panel $b$ shows the CMD for the {\em bulges}. Bulge colors are consistently red regardless of $B/T$ and bulge luminosity. The simplest explanation for this is that bulges are essentially of similar stellar populations, although the age-metallicity degeneracy on optical colors \citep{wort94, yi03, john12, john14} must also be remembered. If we take the median values (as shown in this figure) bulges are slightly redder than the integrated light of the galaxies (panel $a$). This is expected as decomposition removes the (generally blue) disk component from the bulge light in most galaxies. 

There is a larger scatter in color among the bulges in the lower $B/T$ galaxies. In the lowest $B/T$ bin, the extremely red bulge colors (red hatched region on the top) can be interpreted in the same manner as discussed in Section \ref{sec:Optical Colors of Bulges} (unphysical values originated from the shortcoming of the decomposition technique). The scatter on the blue side (blue hatched region in the bottom) however appears to have physical origins. Visual inspection of their optical images suggests the presence of blue star forming regions. 

When we check the emission line properties of the galaxies within the blue dash-line box using the SDSS-fiber spectroscopic data, 88\% (491 galaxies out of 557 galaxies) of them are classified as actively star forming galaxies based on the diagnostics proposed by \citet[]{bald81} (e.g. the BPT diagram). For comparison, only 44\% of the redder ($0.65 < g-r < 1.2$) bulges in the same $B/T$ bin are classified as star forming galaxies. Admittedly, the 3'' diameter SDSS fiber collects light from both bulge and disk components, but it is still useful to check the relative significance of star formation between red and blue bulges.

%---------------------------Section Scaling Relations--------------------------------
\subsection{Scaling Relations}
\label{the scaling relations}

We present the bulge properties in various scaling relations in this section and compare them with those of elliptical galaxies.

%---------------------------Kormendy Relation--------------------------------
\subsubsection{Kormendy Relation}
\label{the scaling relations: Kormendy relation}
\begin{figure}
\centering
\includegraphics[width=0.5\textwidth]{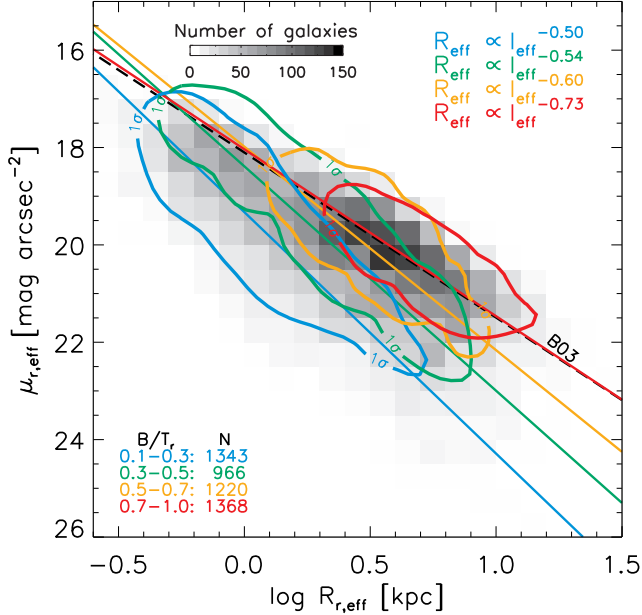}
\caption{The Kormendy relation for {\em{bulges}}. The gray hess diagram shows the whole sample, and the contours show the subsamples binned by $B/T$. The straight lines show the linear fits as explicitly given in the top right corner. The black-dashed line (B03) is the relation for early-type galaxies found by \cite{bern03b}. The bulges in lower \bt galaxies show progressively larger departure from the early-type galaxy sequence.}
\label{fig8}
\end{figure}

The Kormendy scaling relation has been used to study the structural parameters of both ellipticals and spiral bulges \citep{korm77, bern03a, laba03, gado09, fish10}. Specifically, it shows the relation between the effevtive surface brightness and effective radius for the bulges in our sample, binned by $B/T$, as shown in Fig. \ref{fig8}. The fitting coefficients are given in Table \ref{tab3}. The dashed line shows the maximum likelihood of the Kormendy relation suggested for early-type galaxies by \cite{bern03b}. The bulges in the highest $B/T$ galaxies ($0.7 < B/T_{\rm r} \leq 1.0$) closely overlap with the relation derived for early-type galaxies. 
But the departure becomes larger for the bulges in smaller $B/T$ galaxies. This trend implies that the bulges in lower $B/T$ galaxies have a lower stellar density for any given effective radius.

%----------Figure 8---Kormendy relation

``Exponential'' bulges are often considered as a product of secular evolution of bar or disk structure \citep{caro99,korm04}. Specifically, \cite{caro99} reported that the mean surface brightness of their exponential bulges are significantly fainter than those of ``classical'' bulges. The bulges in our low $B/T$ galaxies ($0.1 < B/T_{\rm r} \leq 0.3$) seem to be consistent with their exponential bulges, while most of the bulges in our high $B/T$ ($0.7 < B/T_{\rm r} \leq 1.0$) galaxies seem more comparable to classical bulges. Intermediate $B/T$ bulges lie smoothly in between. This all naturally leads to the possibility that bulges have different formation mechanisms depending on their relative size \citep{silk99,caro99,korm04,fish10,fern14}

%---------------------------Faber-Jackson Relation--------------------------------
\subsubsection{Faber-Jackson Relation}
\label{the scaling relations: Faber-Jackson Relation}

\begin{figure}
\centering
\includegraphics[width=0.5\textwidth]{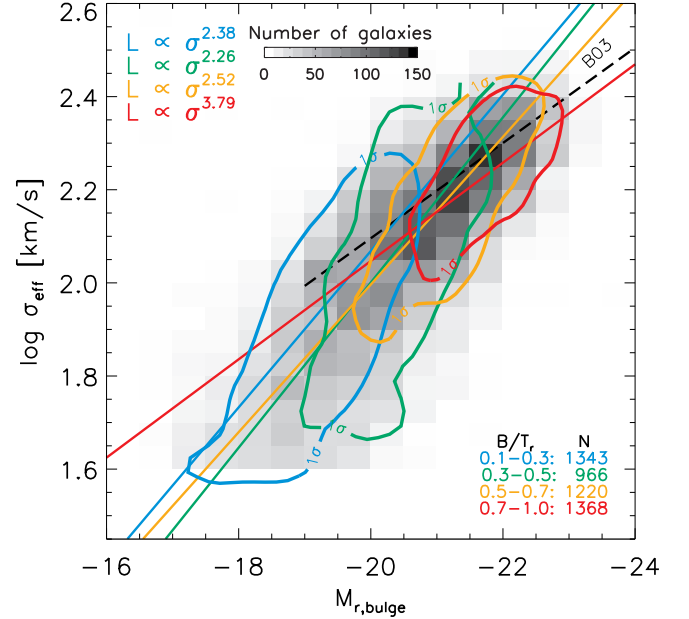}
\caption{The Faber-Jackson relation for {\em{bulges}}. The format is the same as in Fig. \ref{fig8}. A large fraction of the bulges in the low \bt galaxies (e.g., blue contour) show low values of velocity dispersion than found for the bulges in large \bt galaxies, which may hint at kinematics influenced more by ordered motion.}
\label{fig9}
\end{figure}

We now inspect if bulges follow the same Faber-Jackson relation \citep{fabe76} as early-type galaxies. Fig. \ref{fig9} shows our bulges in comparison with the relation found for early-type galaxies by \cite{bern03b}. We fit the bulge properties in this plane ($\rm log\ \sigma$ vs. $M_{\rm r,bulge}$) using a linear fit and transform it into the $L-\sigma$ relation. The fits are given in Table \ref{tab3}. 

%----------Figure 9-----Faber-Jackson relation

When all the bulges are used, we find a reasonably tight correlation but with a slope (3.01) that is much shallower than that of early-type galaxies. The bulges in the highest $B/T$ galaxies (red contour) seem to be in reasonable agreement with early-type galaxies: the slope in our sample is 3.79 and that of early-type galaxies is 3.92. However, the bulges in lower $B/T$ galaxies gradually show larger departure from the early-type galaxy sequence. This trend does not seem to be caused by the inclination effects. We checked the relation between the apparent axis ratio of galaxy and the central velocity dispersion in low $B/T$ galaxies ($0.1 < B/T_{r} \leq 0.3$) and found no significant trend.

The gradual displacement with decreasing $B/T$ from the early-type sequence is clearly visible. Note that many of the bulges in the small $B/T$ galaxies were removed from this diagram because their velocity dispersion was measured to be too small compared to the measurement uncertainty, as mentioned in Section 4.2. If we could measure their velocity dispersions accurately and include them in this analysis, the true trend would likely be even more dramatic. This trend's slope is often interpreted as a result of decreasing mass-to-light ratio \citep[e.g.,][]{korm04}. The bulges in the lowest $B/T$ galaxies have properties consistent with those of pseudobulges \citep[][see their Fig.18]{korm04}. 
The star formation in the bulges of our low $B/T$ galaxies seems to have been persistent, because otherwise a simple fading (by aging) of the stellar population would bring them closer to the early-type sequence quickly, while we do not see such faint bulges in the low velocity dispersion ($\rm log\,\sigma \sim 1.8$) regions. Star formation in larger bulges seems to have ceased earlier. 
This finding is consistent with what we discussed on the Kormendy relation in the previous section.

The departure from the early-type sequence of the bulges for low $B/T$ galaxies can also be explained by differences in kinematic structure. For a given bulge luminosity (e.g., $M_{\rm r,bulge} \sim -18$) lower $B/T$ bulges have a lower velocity dispersion than what is expected from the early-type relation. This is possible if the bulges in the low $B/T$ galaxies are more dominated by (disky) ordered motions \citep{korm82, korm04}. Mergers likely disrupt the previous momentum of merging systems and thus increase the velocity dispersion. Small {\em bulges} will grow in mass and velocity dispersion through mergers, eventually populating in the region that is occupied by the bulges in the highest $B/T$ galaxies, in a similar way suggested for {\em galaxies} \citep{desr07,bern11,mont15}.

%------------------------Table 3;Fitting coefficients of the Kormendy and Faber-Jackson
\begin{table}
\centering
\begin{threeparttable}
\caption{The fitting exponents of Kormendy relation and Faber-Jackson relation in $r$ band}
\label{tab3}    
\begin{tabular}{c c c} \hline \hline

$B/T$ range	&	Kormendy\tnote{a}	&	Faber-Jackson\tnote{b} \\	  
\hline

$0.1 - 0.3$	&  $-0.50 \pm 0.01$	  &   $2.38 \pm 0.28$  \\

$0.3 - 0.5$	&  $-0.54 \pm 0.01$   &   $2.26 \pm 0.31$  \\

$0.5 - 0.7$ &  $-0.60 \pm 0.01$   &   $2.52 \pm 0.22$  \\

$0.7 - 1.0$	&  $-0.73 \pm 0.01$   &   $3.79 \pm 0.47$  \\
\hline
B03\tnote{c} & $-0.75 \pm 0.02$  & $3.91 \pm 0.20$ \\ 
\hline 

\end{tabular}
{\small
\begin{tablenotes}
\item[a] The exponent n of the Kormendy relation: $R_{\rm r,eff} \propto {I_{\rm r,eff}}^{\rm n}.$
\item[b] The exponent n of the Faber-Jackson relation: $L_{\rm r,eff} \propto {\sigma_{\rm eff}}^{\rm n}.$
\item[c] The fitting exponents of the Kormendy and Faber-Jackson relations from \cite{bern03b}, respectively. 
\end{tablenotes}
}
\end{threeparttable}
\end{table}

%---------------------------Section The Fundamental Plane--------------------------------  
\subsubsection{The Fundamental Plane}
\label{the scaling relations: the Fundamental Plane}
 
%----------Figure 10----Fundamental Plane
\begin{figure}
\centering
\includegraphics[width=0.5\textwidth]{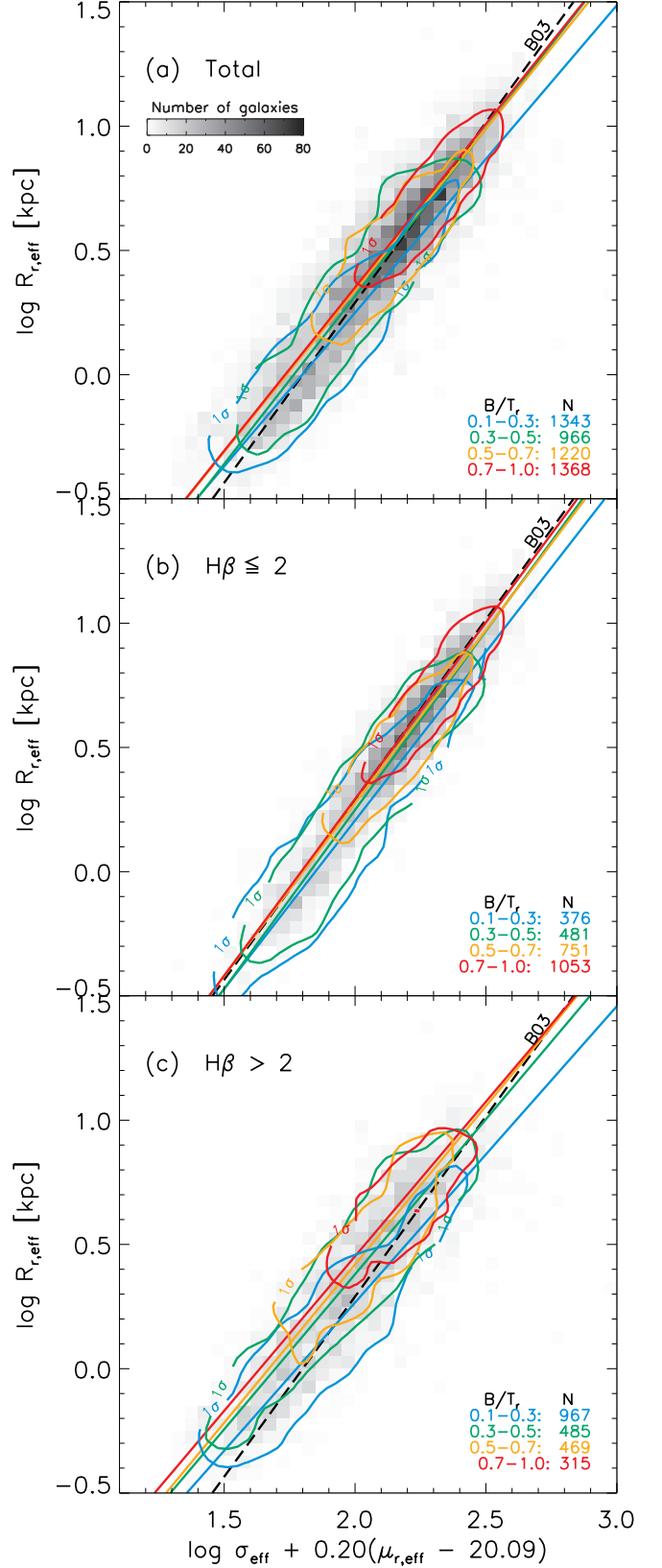}
\caption{The fundamental plane for {\em{bulges}} binned by $B/T$. 
The whole sample ({\em{panel a}}) of bulges shows a FP that is highly close to that of early-type galaxies found by \cite{bern03b}. 
The departure of the slope of the bulges from that of the early-type galaxy reference sample is negligible.
When we divide the sample by $H\beta$, the subsample of $H\beta \leq 2.0$ ({\em{panel b}}) shows a smaller slope departure and RMS scatter (see also Tab.\ref{tab4}) compared with the subsample of $H\beta > 2.0$ ({\em{panel c}}). Note that the a high value of $H\beta$ indicates a smaller SSP age.
}
\label{fig10}
\end{figure}

Elliptical galaxies exhibit a tight sequence in the fundamental plane \citep[][hereafter FP]{dres87,djor87}, and it is generally attributed to the state of dynamical equilibrium and relatively constant stellar properties. We hereby check if bulges follow the same trend. 

The FP of our bulges is shown in Fig. \ref{fig10}. The reference line for SDSS early-type galaxies is adopted from \cite{bern03c}. The fitting coefficients are given in Table \ref{tab4}. Bulges exhibit a FP that is very close to that of early-type galaxies. The bulges in the higher $B/T$ galaxies are naturally larger in size and faster in random motion. The impact of $B/T$ does not appear to be obvious at least in terms of the slope.

%---------------------------------Table 4 Fitting coefficient of FP----------------------------------------------
\begin{table*}
\centering
\begin{threeparttable}
\caption{The fitting coefficients of the edge-on view fundamental plane in $r$ band}
\label{tab4}    
\begin{tabular}{c c c c c c c c c c c c c c c c} \hline \hline

$B/T$ range	&	\multicolumn{5}{c}{Total\tnote{a}}	&  \multicolumn{5}{c}{H$\beta\le2$\tnote{a}} & \multicolumn{5}{c}{H$\beta>2$\tnote{a}} \\	  
\hline
 & $\alpha$ & $\sigma_{\alpha}$ & $\beta$ & $\sigma_{\beta}$ & $\rm RMS_{\rm orth}$\tnote{b} & $\alpha$ & $\sigma_{\alpha}$ & $\beta$ & $\sigma_{\beta}$ & $\rm RMS_{\rm orth}$\tnote{b} & $\alpha$ & $\sigma_{\alpha}$ & $\beta$ & $\sigma_{\beta}$ & $\rm RMS_{\rm orth}$\tnote{b} \\	  
\hline
$0.1 - 0.3$	& 1.24 & 0.02 & -2.22 & 0.04 & 0.10 & 1.36 & 0.04 & -2.51 & 0.08 & 0.07 & 1.19 & 0.02 & -2.12 & 0.04 & 0.10 \\

$0.3 - 0.5$	& 1.35 & 0.02 & -2.38 & 0.05 & 0.09 & 1.44 & 0.03 & -2.63 & 0.05 & 0.07 & 1.25 & 0.03 & -2.12 & 0.06 & 0.09 \\

$0.5 - 0.7$ & 1.30 & 0.02 & -2.26 & 0.05 & 0.08 & 1.39 & 0.03 & -2.51 & 0.06 & 0.06 & 1.27 & 0.04 & -2.13 & 0.08 & 0.09 \\

$0.7 - 1.0$	& 1.31 & 0.02 & -2.28 & 0.05 & 0.07 & 1.42 & 0.02 & -2.55 & 0.05 & 0.06 & 1.24 & 0.06 & -2.04 & 0.13 & 0.09 \\
\hline
B03\tnote{c} & 1.45 & 0.05 & -2.61 & 0.08 & 0.05 & & & & & & & & & & \\
\hline

\end{tabular}
{\small
\begin{tablenotes}
\item[a] The orthogonal fitting slope and intercept (a,b) of the following edge-on view FP: \\
$\rm log \ R_{\rm r,eff} \ = \ (\alpha \pm \sigma_{\alpha}) \,[\rm log \ \sigma_{\rm eff} \ + \ 0.20(\mu_{\rm r,eff} \ - \ 20.09)] \ + \ (\beta \pm \sigma_{\beta}).$
\item[b] The scatter orthogonal to the plane.
\item[c] The orthogonal fitting slope, intercept of \cite{bern03c} for their early-type galaxies in case of $\chi^{2}$ Evolution Selection Effects, and the scatter.
\end{tablenotes}
}
\end{threeparttable}
\end{table*}

We divide the sample by the absorption line strength of $H\beta$ which is known to be sensitive to the presence of young stars \citep{trag00a} and thus widely used as a tracer of recent star formation\citep[e.g.,][]{proc02}. As a simple choice, we use a cut of 2\AA. Stellar populations of $H\beta < 2$ are generally assumed to be dominantly composed of old stars \citep{trag00a}. 

Panels $b$ and $c$ show the subsamples of low and high values of \Hb. The low-\Hb sample shows a slope that is closer to the FP of early-type galaxies, with a smaller RMS scatters (see Table 4). A large fraction (1804 out of 2661, 68\%) of them are in a bulge dominant ($B/T_{\rm r} \leq 0.5$) galaxies. For comparison, this fraction becomes much lower (32\%) in the disk-dominant ($B/T_{\rm r} < 0.5$) galaxies. This strongly suggests that the ``tilt'' and scatter in the FP are at least partially originating from the presence of young stars, or the detailed star formation episodes in the recent history. This conclusion on $bulges$ is qualitatively consistent with what has been suggested earlier on $galaxies$ \citep{forb98,choi09,grav09,jeon09,suh10,spri12}.

%=======================Discussion=======================================
\section{Discussion on spectral line properties}
\label{sec:discussion}

We found in the previous section that bulges belonging to galaxies with smaller $B/T$ show a gradually larger departure from the early-type galaxy scaling relations. A mean age difference, in the sense of younger stellar ages in smaller bulges, appears to explain a large fraction of these trends. We here discuss the likely age difference that can be estimated from the spectroscopic data on our sample galaxies.
  
\begin{figure}
\centering
\includegraphics[width=0.5\textwidth]{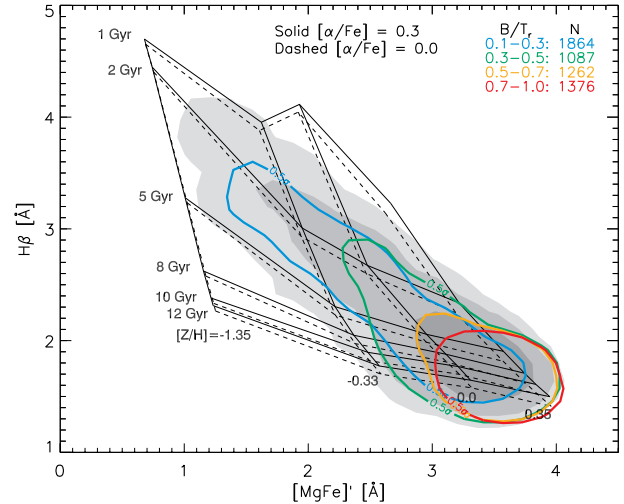}
\caption{The absorption line strengths of galaxies in the 3'' SDSS fiber. 
We adopted the line measurements from the OSSY catalog and applied a cut of signal to statistical noise ratio (S/sN) greater than 10 in all lines.
The gray-shades show the 0.5, 1, and 1.5 $\sigma$ contours and color solid lines show 0.5 $\sigma$ contours for the subsamples binned by $B/T$. 
The SSP model grid has been adopted from \cite{thom03}. 
The solid and dashed grids indicate the models of $[\rm \alpha/ Fe] = 0.3$ and 0.0, respectively.} 
\label{fig11}
\end{figure}

Fig. \ref{fig11} shows the SDSS spectroscopic measurements of our galaxies in comparison to a simple stellar population (SSP) models \citep{thom03}. The SDSS measurements sample the light inside of the 3''-diameter fiber. The median effective diameter of our bulges derived from decomposition is 10.4'', and so the SDSS fiber collects the most of light from the bulge component. Even the smallest bulge in our sample of $B/T_{\rm r} > 0.1$ is 2.4'' in diameter -- only slightly smaller than the SDSS fiber. Among the bulges inspected in our scaling relation studies in Section 4, 90\% of them have $B/T_{\rm r} > 0.6$ within 3'' diameter derived from our decomposition. However, we also note that the fraction reduces down to 70\% in low $B/T$ galaxies ($0.1 < B/T_{\rm r} \leq 0.3$) suggesting that there is some contamination from disk stars in low $B/T$ galaxies. Nonetheless, it is reasonably safe to use the SDSS spectroscopic data to represent bulge properties, except for the low $B/T$ galaxies (see also the discussion of \cite{coel11} on the possible disk contamination in the SDSS fiber.

%----------Figure 11-----MgFe vs Hbeta------

%----------Figure 12---BPT diagram-------
\begin{figure}
\centering
\includegraphics[width=0.5\textwidth]{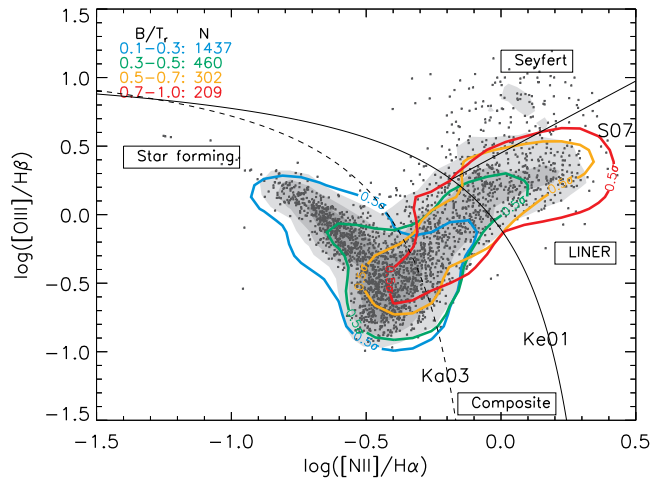}
\caption{The BPT diagnostic diagram. The gray shades are 0.5, 1, and 1.5 $\sigma$ contours for the whole sample. We only use the galaxies that have amplitude over noise (AON) greater than 3 for all four emission lines. The demarcation lines for star formation vs composite/AGN \citep{kewl01}, composite vs AGN \citep{kauf03}, and Seyfert vs LINER \citep{scha07b} are shown. }
\label{fig12}
\end{figure}

The bulges in the highest \bt galaxies are consistent with uniformly old and metal-rich populations, similarly to massive elliptical galaxies. Smaller bulges however show a longer tail toward younger ages and lower metallicity. The $0.5\sigma$ contour of the bulges in the lowest \bt galaxies reaches down to the SSP age of 1\,Gyr. Since there must be a spread in age in real bulge stars, an SSP age of 1\,Gyr effectively means that there currently is star formation.

The low metallicity of the bulges in the low \bt galaxies derived from this analysis is noteworthy. To begin with, it is compatible with the positive relation between galaxy mass and metallicity \citep{trem04,gall05}. Bulges are generally thought be to metal rich, and if this low metallicity derived here for the bulges in the low \bt galaxies is due to the ``frosting effect'' of young stars mixed with a dominant component of old stars, then the youngest population in them would be concluded to be of very low in metallicity. This contradicts the general expectation that younger generations of stars in a closed system are more metal-rich due to chemical recycling. If the low metallicity derived here is true, it strongly implies that bulges are far from being closed systems. Cold gas flows from the vicinity of the baryonic galaxy are probably composed of low metallicity gas which will turn into the formation of low metallicity stars even in the later stage of galaxy evolution. A similar phenomenon has been reported observationally in the Hubble Space Telescope data on the bulge dominant galaxy NGC 4150 \citep{kavi12}.

This phenomenon (younger, yet less metal-rich) is also supported by their emission line properties in the BPT diagram (Fig. \ref{fig12} and Table \ref{tab5}). The bulges in the lower $B/T$ galaxies are more commonly found in the star forming region. The role of a bulge on the star formation and AGN activities of host galaxies has been investigated \citep[e.g.,][]{bell12, sain12, lang14, bluc14a, bluc14b}. The phenomenon seems robust, but it is difficult to conclude solely based on our analysis whether the bulge affects AGN and star formation activities directly, or if it is the other way around.

The decomposition yields disk properties as well as bulge properties. An investigation on disk properties is in progress.

%------------------------Table 5; Fraction of bulges in each BPT region-------------------------------------
\begin{table}
 \begin{center}
 \centering
  \caption [BPT table]
  {Percentage of bulges in different regions in BPT diagram}
   \begin{tabular}{c c c c c}
  \hline \hline
 $B/T$ range &  SF &  composite  & AGN  & Rest \\
 \hline
 $0.1 - 0.3$ &  48.6  & 9.6 & 6.5 & 35.3 \\ 
 $0.3 - 0.5$ &  21.2 & 10.8 & 8.0 & 60 \\
 $0.5 - 0.7$ &  5.9  & 7.3 & 10.5 & 76.3   \\
 $0.7 - 1.0$ &  2.0  & 4.6 &  8.6  & 84.8 \\
 \hline \hline \\
\end{tabular}
\label{tab5}
%\tablenotemark{a}
%\tablenotetext{a}{aaaa}
\tablecomments{The percentages in each line add up to 100\%.}   
\end{center}
\end{table}

%========================Summary======================================
\section{Summary}

To investigate the properties of stellar bulges ranging from the large classical bulges of early-type galaxies to the much smaller bulges of late-type galaxies, we have performed a two-dimensional bulge-disk decomposition for 14,233 galaxies in the redshift range $0.005 < z < 0.05$ from the SDSS DR12 using the program {\sc{galfit}}. The main results and their implications on the bulge formation can be summarized as follows.

It is difficult to derive the size of bulge especially on small-bulge galaxies. This sounds obvious; but in truth it is more a result of the shortcomings of the decomposition technique which affects the colors of the low-\bt galaxies dramatically. As a result, \bt ratios smaller than roughly 0.1 are very uncertain when SDSS data are used. It is important for decomposition studies to explicitly give the details of their decomposition method, as the details have large impacts on the derivation of critical parameters, such as $B/T$. 

For the galaxies of $B/T \geq 0.1$, bulge color is almost constant ($g-r \approx 0.83$) regardless of bulge size. The bulges in the low \bt galaxies show a large spread in $g-r$, and the blue bulges strongly hint at the presence of younger stellar populations.

The results obtained from the three scaling relations (Kormendy relation, Faber-Jackson relation, and the fundamental plane) suggest that bulges have a different mixture of their constituent stellar populations. The bulges in the lower $B/T$ galaxies have fainter surface brightness on average at fixed effective radius in the Kormendy relation, and this variation seems to be gradual as a function of bulge size. 

The bulges in lower \bt galaxies show a lower velocity dispersion for a given bulge luminosity. This hints at the possibility that they are more dominated by (rotational) ordered motion. However, the photometric data we use here are not sufficient to make conclusive comments on kinematic properties of bulges. The stellar {\em light} properties are easier to access through the SDSS data, and thus we investigate on the absorption and emission line properties of effectively ``bulges''. Some bulges in low \bt galaxies appear to be substantially younger than those in large \bt galaxies, roughly by an order of magnitude in SSP age. This is somewhat surprising because we earlier noted that the ``average'' optical $g-r$ color of bulges is roughly fixed against a variation in bulge size. A large fraction of the spread in slope and of the scatter in scaling relations seems to originate from the variation in stellar age, in other words, mass-to-light ratio. A varied mixture of minor and major mergers and a different degree of environmental effect would result in a wide range of \bt which would then naturally explain the smooth distribution of bulges of differing size in the scaling relations.

If the results of our decomposition are robust, most bulges are similar to elliptical galaxies. But it is also clear that the bulges in low \bt galaxies are increasingly deviating from the bulges in early-type galaxies with decreasing $B/T$ ratio. A gradual age difference can explain some of the observed properties but probably does not tell the whole story. If we may combine what is visible in our study with what has been suggested in other studies, it seems natural to conclude that the properties of bulges are a delicate result of the past history of mass build-up including star formation, mergers, and even secular and environmental effects. This may sound obvious because it probably applies to all galaxies. Yet, the variation in the properties of bulges are much wider than that of elliptical galaxies and, accordingly, so maybe their formation processes.

\label{sec:summary}

\acknowledgments
KK has conducted most of the calculations. SO and HJ provided feedback throughout the project. RS and AA-S provided feedback in the step of writing the draft. KK and SKY wrote the paper. H.J. acknowledges support from the Basic Science Research Program through the National Research Foundation of Korea (NRF), funded by the Ministry of Education (NRF-2013R1A6A3A04064993). SKY acknowledges support from the Korean National Research Foundation (NRF-2014R1A2A1A01003730).

%----------------Table 6--------------------------------------------------------------------
\begin{deluxetable*}{ll}
\tablecaption{Galaxy structural parameters from bulge-disk decompositions
\label{tab:tableseven}}
\tabletypesize{\tiny}
\tablewidth{0pt}
\tablehead{ \\ [6pt] \colhead{Parameters}& \colhead{Description}\\}
\startdata
\\ [5pt]
SDSS ObjID & SDSS DR12 Object ID \\ [1pt]
$z_{\rm spec}$ & SDSS Spectroscopic Redshift \\ [1pt]
$Scale_{\rm phy}$ & Conversion factor of arcsec to kpc at redshift z (kpc / arcsec) \\ [1pt]
$R.A._{\rm fit}$ & Fitted R.A. of both bugle and disk (degree) \\ [1pt]
$R.A._{\rm fit,err}$ & Error of fitted R.A. (degree) \\ [1pt]
$Dec._{\rm fit}$ & Fitted Dec. of both of bulge and disk (degree) \\ [1pt]
$Dec._{\rm fit,err}$ & Error of fitted Dec. (degree) \\ [1pt]
$Mag_{\rm g,b}$ & $g$-band apparent magnitude of bulge \\ [1pt]
$Mag_{\rm g,b,err}$ & Error of $g$-band apparent magnitude of bulge \\ [1pt]
$Mag_{\rm r,b}$ & $r$-band apparent magnitude of bulge \\ [1pt]
$Mag_{\rm r,b,err}$ & Error of $r$-band apparent magnitude of bulge \\ [1pt]
$r_{\rm eff}$ & Semimajor effective radius of bulge (pixel) \\ [1pt]
$r_{\rm eff,err}$ & Error of semimajor effective radius of bulge (pixel) \\ [1pt]
$n$ & S\'{e}rsic index of bulge \\ [1pt]
$n_{\rm err}$ & Error of S\'{e}rsic index of bulge \\ [1pt]
$b/a_{\rm b}$ & Axis ratio of bulge \\ [1pt]
$b/a_{\rm b,err}$ & Error of axis ratio of bulge \\ [1pt]
$P.A._{\rm b}$ & Position angle of bulge (degree, from North to East) \\ [1pt]
$P.A._{\rm b,err}$ & Error of position angle of bulge (degree) \\ [1pt]
$Mag_{\rm g,d}$ & $g$-band apparent magnitude of disk \\ [1pt]
$Mag_{\rm g,d,err}$ & Error of $g$-band apparent magnitude of disk \\ [1pt]
$Mag_{\rm r,d}$ & $r$-band apparent magnitude of disk \\ [1pt]
$Mag_{\rm r,d,err}$ & Error of $r$-band apparent magnitude of disk \\ [1pt]
$r_{\rm scl}$ & Semimajor scale length of disk (pixel) \\ [1pt]
$r_{\rm scl,err}$ & Error of semimajor scale length of disk (pixel) \\ [1pt]
$b/a_{\rm d}$ & Axis ratio of disk \\ [1pt]
$b/a_{\rm d,err}$ & Error of axis ratio of disk \\ [1pt]
$P.A._{\rm d}$ & Position Angle of disk in degrees (degree, from North to East) \\ [1pt]
$P.A._{\rm d,err}$ & Error of Position Angle of disk (degree) \\ [1pt]
$ B/T_{\rm g}$ & $g$-band bulge to total ratio \\ [1pt]
$ B/T_{\rm g,err}$ & Error of $g$-band bulge to total ratio \\ [1pt]
$ B/T_{\rm r}$ & $r$-band bulge to total ratio\\ [1pt]
$ B/T_{\rm r,err}$ & Error of $r$-band bulge to total ratio \\ [1pt]
${\chi_{\rm g}}^{2}$ & Reduced $\chi^{2}$ of fit in $g$ band \\ [1pt]
${\chi_{\rm r}}^{2}$ & Reduced $\chi^{2}$ of fit in $r$ band \\ [10pt]

\enddata
\end{deluxetable*}
%\clearpage

%;--------------------Table 7-----------------------------------------------
\begin{deluxetable*}{cccccc}
\tablecolumns{12}
%\tabletypesize{\scriptsize}
\tablecaption{Examples of Galaxy structural parameters \label{tab:tableseven}}
\tablewidth{0pt}
\tablehead{ \\ [7pt]
\colhead{SDSS ObjID} &
\colhead{$z_{\rm spec}$} &
\colhead{$Scale_{\rm phy}$} &
\colhead{$R.A._{\rm fit}$} &
\colhead{$R.A._{\rm fit,err}$} &
\colhead{$Dec._{\rm fit}$} 
}

\startdata
\\ [5pt]
1237655130373161135  &   0.0423376   &   0.83525435   &   242.75532   &   0.00792   &   49.79537   \\ [1pt]
1237654948378837139  &   0.0412290   &   0.81445694   &   243.74803   &   0.00396   &   50.34115   \\ [1pt]
1237654948915839016  &   0.0435533   &   0.85799471   &   244.45464   &   0.00396   &   50.33038   \\ [1pt]
1237668611197698265  &   0.0420984   &   0.83077079   &   244.80742   &   0.00396   &   50.30850   \\ [1pt]
1237654949452709950  &   0.0488313   &   0.95596235   &   245.02977   &   0.00396   &   50.55449   \\ [2pt]
\hline
\\ [1pt]
$Dec._{\rm fit,err}$ & $Mag_{\rm g,b}$ & $Mag_{\rm g,b,err}$ & $Mag_{\rm r,b}$ & $Mag_{\rm r,b,err}$ & $r_{\rm eff}$ \\ [2pt]
\hline
\\ [1pt]
0.00396   &   17.89    &   0.01   &   16.76   &   0.04   &   4.16   \\ [1pt]
0.00396   &   14.97    &   0.00   &   13.94   &   0.01   &   42.93   \\ [1pt]
0.00396   &   15.70    &   0.00   &   14.80   &   0.05   &   18.37   \\ [1pt]
0.00396   &   17.61    &   0.01   &   16.64   &   0.06   &   5.39   \\ [1pt]
0.00396   &   16.55    &   0.00   &   15.69   &   0.05   &   12.25    \\ [2pt]
\hline
\\ [1pt]
$r_{\rm eff,err}$ & $n$ & $n_{\rm err}$ & $b/a_{\rm b}$ & $b/a_{\rm b,err}$ & $P.A._{\rm b}$ \\ [2pt]
\hline
\\ [1pt]
0.32   &   4.94   &   0.42   &   0.30   &   0.01   &   52.84409    \\ [1pt]
1.09   &   5.60   &   0.07   &   0.73   &   0.00   &   223.24847    \\ [1pt]
1.65   &   5.74   &   0.21   &   0.82   &   0.00   &   62.05409    \\ [1pt]
0.50   &   4.48   &   0.25   &   0.49   &   0.01   &    160.41439   \\ [1pt]
0.91   &   3.78   &   0.15   &   0.60   &   0.00   &   126.14346    \\ [2pt]
\hline
\\ [1pt]
$P.A._{\rm b,err}$ & $Mag_{\rm g,d}$ & $Mag_{\rm g,d,err}$ & $Mag_{\rm r,d}$ & $Mag_{\rm r,d,err}$ & $r_{\rm scl}$ \\ [2pt]
\hline
\\ [1pt]
0.54   &    14.41   &   0.00   &   13.81   &   0.00   &   21.70   \\ [1pt]
0.39   &    16.34   &   0.01   &   16.80   &   0.06   &   22.18   \\ [1pt]
0.68   &    16.62   &   0.02   &   16.03   &   0.09   &   32.93   \\ [1pt]
0.52   &    16.43   &   0.00   &   15.79   &   0.02   &   12.66   \\ [1pt]
0.37   &    16.87   &   0.02   &   16.34   &   0.07   &   31.37   \\[2pt]
\hline
\\ [1pt]
$r_{\rm scl,err}$ & $b/a_{\rm d}$ & $b/a_{\rm d,err}$ & $P.A._{\rm d}$ & $P.A._{\rm d,err}$ & $B/T_{\rm g}$ \\ [2pt]
\hline
\\ [1pt]
0.05   &   0.71   &  0.00  & 214.86409 &  0.22 &  0.03897 \\ [1pt]
1.12   &   0.60   &  0.02  & 197.39847 &  2.78 &  0.77934 \\ [1pt]
1.02   &   0.94   &  0.03  & 83.27409 &  16.71 &  0.70001 \\ [1pt]
0.12   &   0.78   &  0.01  & 133.82439 &  1.82 &  0.25222 \\ [1pt]
1.12   &   0.60   &  0.01  & 103.73346 &  2.52 &  0.57315 \\[2pt]
\hline
\\ [1pt]
$B/T_{\rm g,err}$ & $B/T_{\rm r}$ & $B/T_{\rm r,err}$ & ${\chi_{\rm g}}^{2}$ & ${\chi_{\rm r}}^{2}$ \\ [2pt]
\hline
\\ [1pt]
0.00034   &   0.06197    &   0.00214   &   1.702   & 1.680  \\ [1pt]
0.00158   &   0.93303    &   0.00350   &   0.980   & 0.953 \\ [1pt]
0.00387   &   0.75637    &   0.01747   &   0.864   & 0.814 \\ [1pt]
0.00174   &   0.31370    &   0.01254   &   0.999   & 1.023 \\ [1pt]
0.00451   &   0.64535    &   0.01813   &   0.948   & 0.976 \\ [5pt]

\enddata
\tablecomments{Table 7 is published in its entirety in the electronic edition of the Astrophysical Journal Supplement Series. A portion is shown here for guidance regarding its form and content.}
\end{deluxetable*}
\clearpage

%==============================References===========================================================

\clearpage

\begin{thebibliography}{}

\bibitem[\protect\citeauthoryear{Alam et al.}{2015}]{alam15} Alam, S., Albareti, F.~D., 
Allende Prieto, C., et al.\ 2015, \apjs, 219, 12 

\bibitem[\protect\citeauthoryear{Allen et al.}{2006}]{alle06} Allen, P.~D., Driver, 
S.~P., Graham, A.~W., et al.\ 2006, \mnras, 371, 2 

\bibitem[\protect\citeauthoryear{Athanassoula}{2016}]{atha16} Athanassoula, E.\ 2016, 
Galactic Bulges, 418, 391 

\bibitem[\protect\citeauthoryear{Baldwin et al.}{1981}]{bald81}
 Baldwin, J. A., Phillips, M. M., \& Terlevich, R. 1981, \pasp, 93, 5

\bibitem[\protect\citeauthoryear{Baum}{1959}]{baum59} Baum, W.~A.\ 1959, \pasp, 71, 106 

\bibitem[\protect\citeauthoryear{Bell et al.}{2012}]{bell12} Bell, E.~F., van der Wel, 
A., Papovich, C., et al.\ 2012, \apj, 753, 167 

\bibitem[\protect\citeauthoryear{Bender et al.}{1993}]{bend93} Bender, R., Burstein, 
D., \& Faber, S.~M.\ 1993, \apj, 411, 153

\bibitem[\protect\citeauthoryear{Benson et al.}{2007}]{bens07} Benson, A.~J., D{\v 
z}anovi{\'c}, D., Frenk, C.~S., \& Sharples, R.\ 2007, \mnras, 379, 841 

\bibitem[\protect\citeauthoryear{Bernardi et al.}{2003a}]{bern03a}
Bernardi, M., Sheth, R. K., Annis, J., et al. 2003, \aj, 125, 1849

\bibitem[\protect\citeauthoryear{Bernardi et al.}{2003b}]{bern03b} Bernardi, M., Sheth, 
R.~K., Annis, J., et al.\ 2003, \aj, 125, 1882 

\bibitem[\protect\citeauthoryear{Bernardi et al.}{2003c}]{bern03c} Bernardi, M., Sheth, 
R.~K., Annis, J., et al.\ 2003, \aj, 125, 1866 

\bibitem[\protect\citeauthoryear{Bernardi et al.}{2011}]{bern11} Bernardi, M., Roche, 
N., Shankar, F., \& Sheth, R.~K.\ 2011, \mnras, 412, L6 

\bibitem[\protect\citeauthoryear{Bertin \& Arnouts}{1996}]{bert96} Bertin, E., \& Arnouts, S.\ 1996, \aaps, 117, 393 

\bibitem[\protect\citeauthoryear{Blanton et al.}{2003}]{blan03} Blanton, M.~R., Hogg, 
D.~W., Bahcall, N.~A., et al.\ 2003, \apj, 594, 186 

\bibitem[\protect\citeauthoryear{Bluck et al.}{2014a}]{bluc14a} Bluck, A.~F.~L., Mendel, 
J.~T., Ellison, S.~L., et al.\ 2014, \mnras, 441, 599 

\bibitem[\protect\citeauthoryear{Bluck et al.}{2014b}]{bluc14b} Bluck, A.~F.~L., Ellison, 
S.~L., Patton, D.~R., et al.\ 2014, arXiv:1412.3862 

\bibitem[\protect\citeauthoryear{Bower et al.}{1992}]{bowe92b} Bower, R.~G., Lucey, 
J.~R., \& Ellis, R.~S.\ 1992, \mnras, 254, 601 

\bibitem[\protect\citeauthoryear{Broeils 
\& Courteau}{1997}]{broe97} Broeils, A.~H., \& Courteau, S.\ 1997, Dark and Visible Matter in Galaxies and Cosmological Implications, 117, 74 

\bibitem[\protect\citeauthoryear{Byun 
\& Freeman}{1995}]{byun95} Byun, Y.~I., \& Freeman, K.~C.\ 1995, \apj, 448, 563 

\bibitem[\protect\citeauthoryear{Capaccioli}{1989}]{capa89} Capaccioli, M.\ 1989, World 
of Galaxies (Le Monde des Galaxies), 208 

\bibitem[\protect\citeauthoryear{Cappellari \& Emsellem}{2004}]{capp04} 
Cappellari, M., \& Emsellem, E.\ 2004, \pasp, 116, 138 

\bibitem[\protect\citeauthoryear{Cappellari et al.}{2006}]{capp06} Cappellari, M., 
Bacon, R., Bureau, M., et al.\ 2006, \mnras, 366, 1126 

\bibitem[\protect\citeauthoryear{Carollo}{1999}]{caro99} Carollo, C.~M.\ 1999, \apj, 
523, 566 

\bibitem[\protect\citeauthoryear{Choi et al.}{2009}]{choi09} Choi, Y., Goto, T., 
\& Yoon, S.-J.\ 2009, \mnras, 395, 637 

\bibitem[\protect\citeauthoryear{Coelho 
\& Gadotti}{2011}]{coel11} Coelho, P., \& Gadotti, D.~A.\ 2011, \apjl, 743, L13 

\bibitem[\protect\citeauthoryear{Cowie et al.}{1996}]{cowi96} Cowie, L.~L., Songaila, 
A., Hu, E.~M., \& Cohen, J.~G.\ 1996, \aj, 112, 839 

\bibitem[\protect\citeauthoryear{De Lucia et al.}{2006}]{delu06} De Lucia, G., 
Springel, V., White, S.~D.~M., Croton, D., 
\& Kauffmann, G.\ 2006, \mnras, 366, 499 

\bibitem[\protect\citeauthoryear{Desroches et al.}{2007}]{desr07} Desroches, L.-B., 
Quataert, E., Ma, C.-P., \& West, A.~A.\ 2007, \mnras, 377, 402 

\bibitem[\protect\citeauthoryear{Djorgovski 
\& Davis}{1987}]{djor87} Djorgovski, S., \& Davis, M.\ 1987, \apj, 313, 59

\bibitem[\protect\citeauthoryear{Dressler}{1987}]{dres87} Dressler, A.\ 1987, \apj, 
317, 1 

\bibitem[\protect\citeauthoryear{Driver et al.}{2006}]{driv06} Driver, S.~P., Allen, 
P.~D., Graham, A.~W., et al.\ 2006, \mnras, 368, 414 

\bibitem[\protect\citeauthoryear{Erwin et al.}{2015}]{erwi15} Erwin, P., Saglia, R.~P., 
Fabricius, M., et al.\ 2015, \mnras, 446, 4039 


\bibitem[\protect\citeauthoryear{Faber 
\& Jackson}{1976}]{fabe76} Faber, S.~M., \& Jackson, R.~E.\ 1976, \apj, 204, 668 

\bibitem[\protect\citeauthoryear{Fabricius et al.}{2012}]{fabr12} Fabricius, M.~H., 
Saglia, R.~P., Fisher, D.~B., et al.\ 2012, \apj, 754, 67 

\bibitem[\protect\citeauthoryear{Falc{\'o}n-Barroso et al.}{2002}]{falc02} 
Falc{\'o}n-Barroso, J., Peletier, R.~F., 
\& Balcells, M.\ 2002, \mnras, 335, 741 

\bibitem[\protect\citeauthoryear{Fern{\'a}ndez Lorenzo et al.}{2014}]{fern14} 
Fern{\'a}ndez Lorenzo, M., Sulentic, J., Verdes-Montenegro, L., et al.\ 
2014, \apjl, 788, LL39 

\bibitem[\protect\citeauthoryear{Fisher et al.}{1996}]{fish96}
Fisher D., Franx M., Illingworth G., 1996, \apj, 459, 110

\bibitem[\protect\citeauthoryear{Fisher 
\& Drory}{2010}]{fish10} Fisher, D.~B., \& Drory, N.\ 2010, \apj, 716, 942 

\bibitem[\protect\citeauthoryear{Forbes et al.}{1998}]{forb98} Forbes, D.~A., Ponman, 
T.~J., \& Brown, R.~J.~N.\ 1998, \apjl, 508, L43 

\bibitem[\protect\citeauthoryear{Freeman}{1970}]{free70}
Freeman, K.~C.\ 1970, \apj, 160, 811 

\bibitem[\protect\citeauthoryear{Gallazzi et al.}{2005}]{gall05} Gallazzi, A., Charlot, S., Brinchmann, J., White, S.~D.~M., \& Tremonti, C.~A.\ 2005, \mnras, 362, 41 

\bibitem[\protect\citeauthoryear{Gadotti}{2009}]{gado09} Gadotti, D.~A.\ 2009, \mnras, 
393, 1531 

% \bibitem[\protect\citeauthoryear{Graham et al.}{2005}]{grah05} Graham, A.~W., Driver, 
%S.~P., Petrosian, V., et al.\ 2005, \aj, 130, 1535 

\bibitem[\protect\citeauthoryear{Graves et al.}{2009}]{grav09} Graves, G.~J., Faber, 
S.~M., \& Schiavon, R.~P.\ 2009, \apj, 698, 1590 

\bibitem[\protect\citeauthoryear{H{\"a}ussler et al.}{2007}]{haus07} H{\"a}ussler, B., 
McIntosh, D.~H., Barden, M., et al.\ 2007, \apjs, 172, 615 

\bibitem[\protect\citeauthoryear{Hubble}{1926}]{hubb26} Hubble, E.~P.\ 1926, \apj, 64, 
321 

\bibitem[\protect\citeauthoryear{Hubble}{1936}]{hubb36} Hubble, E.~P.\ 1936, The Realm of 
the Nebulae, Yale Univ. Press, Yale.

\bibitem[\protect\citeauthoryear{Hudson et al.}{2010}]{huds10} Hudson, M.~J., 
Stevenson, J.~B., Smith, R.~J., et al.\ 2010, \mnras, 409, 405

\bibitem[\protect\citeauthoryear{Jablonka et al.}{1996}]{jabl96} Jablonka, P., Martin, 
P., \& Arimoto, N.\ 1996, \aj, 112, 1415 

\bibitem[\protect\citeauthoryear{Jeong et al.}{2009}]{jeon09} Jeong, H., Yi, S.~K., 
Bureau, M., et al.\ 2009, \mnras, 398, 2028 

\bibitem[\protect\citeauthoryear{Johnston et al.}{2012}]{john12} Johnston, E.~J., Arag{\'o}n-Salamanca, A., Merrifield, M.~R., \& Bedregal, A.~G.\ 2012, \mnras, 422, 2590 

\bibitem[\protect\citeauthoryear{Johnston et al.}{2014}]{john14} Johnston, E.~J., Arag{\'o}n-Salamanca, A., \& Merrifield, M.~R.\ 2014, \mnras, 441, 333

\bibitem[\protect\citeauthoryear{Kaviraj et al.}{2012}]{kavi12} Kaviraj, S., Crockett, 
R.~M., Whitmore, B.~C., et al.\ 2012, \mnras, 422, L96 

\bibitem[\protect\citeauthoryear{Kauffmann et al.}{2003}]{kauf03} Kauffmann, G., 
Heckman, T.~M., Tremonti, C., et al.\ 2003, \mnras, 346, 1055 

\bibitem[\protect\citeauthoryear{Kelvin et al.}{2012}]{kelv12}
Kelvin, L.~S., Driver, S.~P., Robotham, A.~S.~G., et al.\ 2012, \mnras, 421, 1007 

\bibitem[\protect\citeauthoryear{Kent}{1985}]{kent85} Kent, S.~M.\ 1985, \apjs, 59, 115 

\bibitem[\protect\citeauthoryear{Kewley et al.}{2001}]{kewl01} Kewley, L.~J., Dopita, 
M.~A., Sutherland, R.~S., Heisler, C.~A., 
\& Trevena, J.\ 2001, \apj, 556, 121 

\bibitem[\protect\citeauthoryear{Khim et al.}{2015}]{khim15} Khim, H.-g., Park, J., 
Seo, S.-W., et al.\ 2015, arXiv:1507.04069 

\bibitem[\protect\citeauthoryear{Kim et al.}{2008}]{kim08} Kim, M., Ho, L.~C., Peng, 
C.~Y., et al.\ 2008, \apj, 687, 767 

\bibitem[\protect\citeauthoryear{Kormendy}{1977}]{korm77} Kormendy, J.\ 1977, \apj, 
218, 333 

\bibitem[\protect\citeauthoryear{Kormendy 
\& Illingworth}{1982}]{korm82} Kormendy, J., \& Illingworth, G.\ 1982, \apj, 256, 460 

\bibitem[\protect\citeauthoryear{Kormendy \& Kennicutt}{2004}]{korm04}
Kormendy, J., \& Kennicutt, R.~C., Jr.\ 2004, \araa, 42, 603 

\bibitem[\protect\citeauthoryear{Kormendy 
\& Ho}{2013}]{korm13} Kormendy, J., \& Ho, L.~C.\ 2013, \araa, 51, 511 

\bibitem[\protect\citeauthoryear{Kormendy}{2015}]{korm15} Kormendy, J.\ 2015, 
arXiv:1504.03330 

\bibitem[\protect\citeauthoryear{La Barbera et al.}{2003}]{laba03} La Barbera, F., 
Busarello, G., Merluzzi, P., Massarotti, M., 
\& Capaccioli, M.\ 2003, \apj, 595, 127 

\bibitem[\protect\citeauthoryear{Lackner \& Gunn}{2012}]{lack12}
Lackner, C.~N., \& Gunn, J.~E.\ 2012, \mnras, 421, 2277 

\bibitem[\protect\citeauthoryear{Lang et al.}{2014}]{lang14} Lang, P., Wuyts, S., 
Somerville, R.~S., et al.\ 2014, \apj, 788, 11 

\bibitem[\protect\citeauthoryear{Lee 
\& Yi}{2013}]{lee13} Lee, J., \& Yi, S.~K.\ 2013, \apj, 766, 38 

\bibitem[\protect\citeauthoryear{MacArthur et al.}{2003}]{maca03} MacArthur, L.~A., 
Courteau, S., \& Holtzman, J.~A.\ 2003, \apj, 582, 689 

\bibitem[\protect\citeauthoryear{Meert et al.}{2013}]{meer13}
Meert, A., Vikram, V., \& Bernardi, M.\ 2013, \mnras, 433, 1344 

\bibitem[\protect\citeauthoryear{Meert et al.}{2015}]{meer15}
Meert, A., Vikram, V., \& Bernardi, M.\ 2015, \mnras, 446, 3943 

\bibitem[\protect\citeauthoryear{Mendel et al.}{2014}]{mend14} Mendel, J.~T., Simard, 
L., Palmer, M., Ellison, S.~L., \& Patton, D.~R.\ 2014, \apjs, 210, 3

\bibitem[\protect\citeauthoryear{Montero-Dorta et al.}{2015}]{mont15} Montero-Dorta, 
A.~D., Shu, Y., Bolton, A.~S., Brownstein, J.~R., 
\& Weiner, B.~J.\ 2015, arXiv:1505.03866 

\bibitem[\protect\citeauthoryear{Oh et al.}{2011}]{oh11}                                                      
Oh, K., Sarzi, M., Schawinski, K., \& Yi, S. K. 2011, \apjs, 195, 13

\bibitem[\protect\citeauthoryear{Oh et al.}{2012}]{oh12} Oh, S., Oh, K., 
\& Yi, S.~K.\ 2012, \apjs, 198, 4 

\bibitem[\protect\citeauthoryear{Oh et al.}{2013}]{oh13} Oh, K., Choi, H., Kim, 
H.-G., Moon, J.-S., \& Yi, S.~K.\ 2013, \aj, 146, 151 

\bibitem[\protect\citeauthoryear{Peletier et al.}{1999}]{pele99} Peletier, R.~F., 
Balcells, M., Davies, R.~L., et al.\ 1999, \mnras, 310, 703 

\bibitem[\protect\citeauthoryear{Peng et al.}{2002}]{peng02} Peng, C.~Y., Ho, L.~C., 
Impey, C.~D., \& Rix, H.-W.\ 2002, \aj, 124, 266 

\bibitem[\protect\citeauthoryear{Peng et al.}{2010}]{peng10}
Peng, C.~Y., Ho, L.~C., Impey, C.~D., \& Rix, H.-W.\ 2010, \aj, 139, 2097 

\bibitem[\protect\citeauthoryear{Proctor 
\& Sansom}{2002}]{proc02} Proctor, R.~N., \& Sansom, A.~E.\ 2002, \mnras, 333, 517

\bibitem[\protect\citeauthoryear{Saintonge et al.}{2012}]{sain12} Saintonge, A., 
Tacconi, L.~J., Fabello, S., et al.\ 2012, \apj, 758, 73 

\bibitem[\protect\citeauthoryear{Sandage 
\& Visvanathan}{1978}]{sand78} Sandage, A., \& Visvanathan, N.\ 1978, \apj, 223, 707 

\bibitem[\protect\citeauthoryear{Sarzi et al.}{2006}]{sarz06}
Sarzi, M., et al.\ 2006, \mnras, 366, 1151 

\bibitem[\protect\citeauthoryear{Schawinski et al.}{2007}]{scha07b} Schawinski, K., 
Thomas, D., Sarzi, M., et al.\ 2007, \mnras, 382, 1415 

\bibitem[\protect\citeauthoryear{S\'{e}rsic}{1968}]{sers68}
S\'{e}rsic, J. L. 1968, Atlas de Galaxies Australes (C\'{o}rdoba: ObS, Astron., Univ. Nac. C\'{o}rdoba 

\bibitem[\protect\citeauthoryear{Silk \& Bouwens}{1999}]{silk99} Silk, J., \& Bouwens, R.\ 1999, \apss, 265, 379 

\bibitem[\protect\citeauthoryear{Simard et al.}{2002}]{sima02} Simard, L., Willmer, 
C.~N.~A., Vogt, N.~P., et al.\ 2002, \apjs, 142, 1 

\bibitem[\protect\citeauthoryear{Simard et al.}{2011}]{sima11}
Simard, L., Mendel, J.~T., Patton, D.~R., Ellison, S.~L., \& McConnachie, A.~W.\ 2011, \apjs, 196, 11 

\bibitem[\protect\citeauthoryear{Simien 
\& de Vaucouleurs}{1986}]{simi86} Simien, F., \& de Vaucouleurs, G.\ 1986, \apj, 302, 564 

\bibitem[\protect\citeauthoryear{Springob et al.}{2012}]{spri12} Springob, C.~M., 
Magoulas, C., Proctor, R., et al.\ 2012, \mnras, 420, 2773 

\bibitem[\protect\citeauthoryear{Strateva et al.}{2001}]{stra01} Strateva, I., 
Ivezi{\'c}, {\v Z}., Knapp, G.~R., et al.\ 2001, \aj, 122, 1861 

\bibitem[\protect\citeauthoryear{Suh et al.}{2010}]{suh10} Suh, H., Jeong, H., Oh, K., 
et al.\ 2010, \apjs, 187, 374 

\bibitem[\protect\citeauthoryear{Thomas et al.}{2003}]{thom03} Thomas, D., Maraston, 
C., \& Bender, R.\ 2003, \mnras, 339, 897 

\bibitem[\protect\citeauthoryear{Toomre 
\& Toomre}{1972}]{toom72} Toomre, A., \& Toomre, J.\ 1972, \apj, 178, 623 

\bibitem[\protect\citeauthoryear{Trager et al.}{2000}]{trag00a} Trager, S.~C., Faber, 
S.~M., Worthey, G., \& Gonz{\'a}lez, J.~J.\ 2000, \aj, 119, 1645 

\bibitem[\protect\citeauthoryear{Tremonti et al.}{2004}]{trem04} Tremonti, C.~A., 
Heckman, T.~M., Kauffmann, G., et al.\ 2004, \apj, 613, 898 

\bibitem[\protect\citeauthoryear{Weinzirl et al.}{2009}]{wein09} Weinzirl, T., Jogee, 
S., Khochfar, S., Burkert, A., \& Kormendy, J.\ 2009, \apj, 696, 411 

\bibitem[\protect\citeauthoryear{White 
\& Rees}{1978}]{whit78} White, S.~D.~M., \& Rees, M.~J.\ 1978, \mnras, 183, 341 

\bibitem[\protect\citeauthoryear{Worthey}{1994}]{wort94} Worthey, G.\ 1994, \apjs, 95, 107
 
 \bibitem[\protect\citeauthoryear{Yi}{2003}]{yi03} Yi, S.~K.\ 2003, \apj, 582, 202

\bibitem[\protect\citeauthoryear{York et al.}{2000}]{york00}
York, D. G., et al. 2000, \aj, 120, 1579

\bibitem[\protect\citeauthoryear{Zhao et al.}{2015}]{zhao15} Zhao, D., 
Arag{\'o}n-Salamanca, A., \& Conselice, C.~J.\ 2015, \mnras, 448, 2530 

\end{thebibliography}
\end{document}